\documentclass[seceq]{ptptex}
\usepackage{wrapft}

\newcommand{\lap}[0]{\int^{-\infty}_{0} dp \hspace*{1mm} e^{-pz} \hspace*{1mm}}
\newcommand{\sig}[1]{\sum^{\infty}_{#1=0}}

\newcommand{\sh}[1]{\sinh\sqrt{k}#1}
\newcommand{\ch}[1]{\cosh\sqrt{k}#1}

\usepackage{amssymb}
\usepackage[dvips]{graphicx}
\usepackage{booktabs}
\usepackage{tabularx}
\newcommand{\pmat}[1]{\begin{pmatrix}#1\end{pmatrix}}
\usepackage{graphicx,color}

\title{Reconnection of Unstable/Stable Manifolds of the Harper Map 
}

\subtitle{ {\large
Asymptotics-Beyond-All-Orders Approach} }    

\author{
Shigeru \textsc{Ajisaka}\footnote{Email: g00k0056@suou.waseda.jp}
 and Shuichi \textsc{Tasaki}
}

\inst{
Advanced Institute for Complex Systems
and 
Department of Applied Physics,\\
School
of Science and Engineerings,
Waseda University,\\
3-4-1 Okubo, Shinjuku-ku,
Tokyo 169-8555,
Japan
}

\abst{
The Harper map is one of the simplest chaotic systems exhibiting 
reconnection of invariant manifolds. 
The method of asymptotics beyond all orders (ABAO) is used to construct 
unstable/stable manifolds of the Harper map.
By enlarging the neighborhood of a singularity, the perturbative solution of the unstable manifold is 
expressed as a Borel summable asymptotic expansion in a sector including 
$t=-\infty$ and is analytically continued to the other sectors, where 
the solution acquires new terms describing heteroclinic tangles.
When the parameter changes to the reconnection threshold, the 
unstable/stable manifolds are shown to acquire new oscillatory portion 
corresponding to the heteroclinic tangle after the reconnection. 
}

\begin{document}
\maketitle

\section{Introduction}

Bifurcation involving chaotic motion is one of the origins of 
diversity in complex systems and reconnection among unstable/stable 
manifolds is such an example. 
However, general features of the interplay between bifurcation 
and chaos are not well understood and a case study on a simple system
would provide useful information for the generic behavior. 
Moreover, since the trajectory near the bifurcation point is considered 
to be very complicated, an analytical approach would be
more helpful. Thus, as one of the simplest systems exhibiting 
reconnection,\cite{Saito,Shinohara} 
we analytically study the nearly integrable 
Harper map.

\begin{figure}[b]
\label{fig:differential}
\begin{center}
\includegraphics[width=12cm,keepaspectratio,clip]{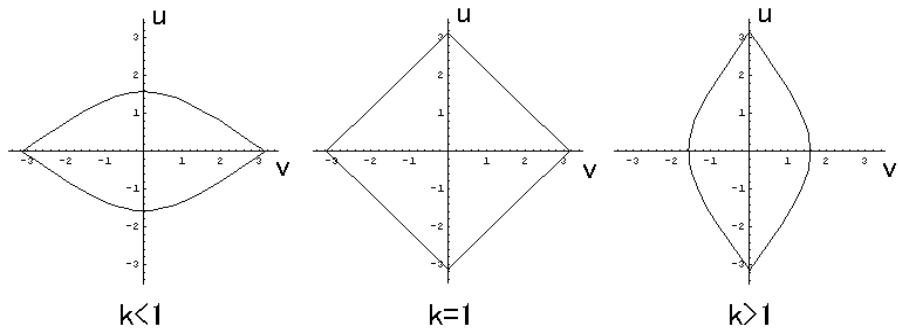}
\caption{Separatrix of the Harper map in the continuous limit.}
\end{center}
\end{figure}

The Harper map depends on a real parameter $k$ and is defined 
on the torus\\$\{(v,u)~\in~[-\pi,\pi]^2\}$:
\begin{eqnarray}
v(t+\sigma)-v(t)&=&-\sigma \sin u(t)
\nonumber \\
u(t+\sigma)-u(t)&=&k\sigma \sin v(t+\sigma)
\label{eq:Harper}
\end{eqnarray}
where 
$\sigma (>0)$ is the time step and it plays a role of the small
parameter. 
Since the case of $k<0$ is conjugate to that of $k>0$~\cite{Shinohara}
and since the solution for $k>1$ can be obtained from that for $k<1$ by a simple 
symmetry argument (cf. Appendix \ref{appC}),  it is sufficient to consider the case of 
$0<k\le 1$.
In the continuous time limit, where $\sigma\to 0$, the map
reduces to an integrable set of differential equations:
\begin{eqnarray}
v'(t)&=&- \sin u(t) \nonumber
\\
u'(t)&=&k \sin v(t) \ . 
\label{eq:Harper0}
\end{eqnarray}
Throughout this paper, the prime is used to indicate the differentiation
with respect to time $t$ such as $u'(t)\equiv{\displaystyle du(t)\over
\displaystyle dt}$.
Eq.~(\ref{eq:Harper0}) admits topologically different separatrices 
depending on the parameter $k$ (cf.~Fig.~\ref{fig:differential})
and the separatrix changes its shape when $k\to 1$.
As shown in the figure, when $k$ increases, the middle point of the upper 
separatrix moves upward and, when $k=1$, it reaches the fixed point $(0,\pi)$.
This view would be changed for nonvanishing $\sigma$ due to the heteroclinic 
tangles of stable and unstable manifolds. Indeed, in case of $0<k<1$, 
the unstable manifold of the fixed point $(\pi,0)$
is expected to show an oscillatory behavior near $(-\pi,0)$, while 
in case of $k=1$, it would oscillate near $(-\pi,0)$ 
{\it and} $(0,\pi)$.
We, then, study the crossover between the two behaviors
based on an analytical solution.

One of the powerful methods analytically dealing with heteroclinic/homoclinic 
tangles of invariant manifolds is the method of asymptotics beyond all 
orders (ABAO method). The key idea is to employ the so-called inner 
equation and to investigate it with the aid of 
the resurgence theory\cite{Ecall} and the Borel 
resummation. The inner equation magnifies the behavior of the solution 
near its singularities and bridges the solutions in different sectors.

The idea of the inner equation was first 
used by Lazutkin\cite{Lazutkinf}
to derive the first crossing angle between the stable and unstable 
manifolds of Chirikov's standard map, and by Kruskal and Segur\cite{Kruskal}, 
to study a singular perturbation problem of ordinary differential equations.
Lazutkin and his 
coworkers\cite{Lazutkin,Gelfreich5,Lazutkin91,Lazutkin92} developed their 
method, and Lazutkin, Gelfreich and Svanidze\cite{GLS94} proposed a method 
of systematically improving the solution of the inner equation for
the standard map. The first rigorous proof of the method was given 
by Gelfreich.~\cite{Gelfreich96,Gelfreich99}
He and his collaborators\cite{Gelfreich1,Gelfreich2,Gelfreich3,Gelfreich4} 
have also studied the splitting of separatrices and related bifurcations 
for various systems.
Meanwhile, 
Hakim and Mallick\cite{Hakim} 
used
the technique of the Borel transformation within this context and obtained 
the first crossing angle previously obtained in Ref.~\citen{Lazutkinf}.
Tovbis, Tsuchiya and Jaff\'e~\cite{Tovbis} improved their method and
derived analytical approximations of perturbed unstable/stable 
manifolds for the H\'enon map.
Nakamura and Hamada~\cite{NakamuraHamada,Nakamurabook} studied the 
separatrix splitting
for the cubic map and,
as in the work of Ref.~\citen{GLS94},
Nakamura and Kushibe~\cite{NakamuraKushibe}
considered the systematic improvement of the solution of the standard map.
Also, ABAO method was applied to some higher dimensional systems
by Hirata, Nozaki and Konishi.~\cite{Hirata1,Hirata2}

Roughly speaking, the procedure obtaining an approximate expression of
the unstable manifold can be summarized as follows:
\begin{enumerate}

\item[(i)] Find singularities in the complex time domain of the lowest 
order perturbative solution.

\item[(ii)] Magnify the neighborhood of a singularity closest to the
real axis and derive an asymptotic expansion of the lowest order
solution valid in a certain sector.

\item[(iii)] The Borel transformation and the resurgence theory are used to
construct the asymptotic expansion which is valid in the other sector.
Usually, there appear additional terms which are exponentially small 
for real time values.

\item[(iv)] Going back to the original equation, derive equations 
for the exponentially small corrections corresponding to the
terms found in the previous step. An appropriate solution is chosen 
by matching its asymptotic expansion with that obtained in 
the previous step.

\end{enumerate}
In this paper, with the aid of the ABAO method, the reconnection of the
unstable manifold is studied for the Harper map. 
Our presentation mainly follows the work of 
Nakamura and Kushibe~\cite{NakamuraKushibe}. 
In the text, an approximate expression of the unstable manifold
is constructed and only the result is given for the stable manifold.

The rest of this paper is arranged as follows: In the next section,
an outline of our result is presented without proofs and 
the details, which are given in Secs.~3-6.
In Sec.~3, a perturbative solution is constructed by the Melnikov 
perturbation method and its asymptotic expansion is derived.  
In Sec.~4, the resurgence theory and the Borel resummation method are 
applied to solve the inner equation, and asymptotic expansions of
the additional terms are derived.
In Sec.~5, the exponentially small corrections are obtained by
matching the solutions of the original and inner equations.
As will be discussed later, the separatrix solution of (\ref{eq:Harper0}) has 
singularities along two lines both parallel to the imaginary axis 
and, among them two pairs on individual lines mainly contribute to the 
heteroclinic tangles of the manifolds.
Secs.~4 and 5 deal with one pair of singularities which have smaller 
real part and, in Sec.~6, the same analysis is repeated for 
the other pair of singularities. 
In Sec.~7, the derived solution is compared with numerical calculations
and the reconnection of the unstable manifold is discussed.
Sec.~8 is devoted to summary.

\section{Outline of the Construction of Unstable Manifold}

In this section, we summarize the construction of the unstable manifolds 
of the Harper map with the aid of the method of asymptotics beyond all 
orders (ABAO method). The details of the analysis will be described in 
Secs.~\ref{Sec3}-\ref{sec:other}.

\subsection{Unstable manifold}

We are interested in the unstable manifold of the Harper map 
satisfying the boundary condition:
\begin{eqnarray}
v(-\infty)=\pi ,\ \ 
u(-\infty)=0
\label{bc}
\end{eqnarray}
When the parameter $\sigma$ is small enough, the solution of 
(\ref{eq:Harper0}) with the same boundary condition provides 
a good approximation for $t\ll 0$. However, 
the solution is not valid for the whole real $t$ since exponentially 
small terms should be added when $t$ exceeds certain values 
$t=T_1,T_2\ (T_1<T_2)$. In fact, the unstable manifold is 
well-approximated by a double expansion with respect to $\sigma$ 
and $\epsilon\equiv \sigma^j e^{-{c\over \sigma}}$ 
($c>0, j$: some constant):
\begin{eqnarray}
\begin{pmatrix}
v(t) \cr u(t)
\end{pmatrix}
&=&
\begin{pmatrix}
v_0(t,\sigma) \cr u_0(t,\sigma)
\end{pmatrix}
+S(t-T_1)\sum_{n=1}^\infty
\begin{pmatrix}
{\rm Re}\left(E_{1}(t)^n v_n^{(1)}(t,\sigma)\right) \cr
{\rm Re}\left(E_{1}(t)^n u_n^{(1)}(t,\sigma)\right)
\end{pmatrix}
\cr
&&\mskip 80mu +S(t-T_2)\sum_{n=1}^\infty
\begin{pmatrix}
{\rm Re}\left(E_{2}(t)^n v_n^{(2)}(t,\sigma,\epsilon)\right) \cr
{\rm Re}\left(E_{2}(t)^n u_n^{(2)}(t,\sigma,\epsilon)\right)
\end{pmatrix}
\label{2.2st}
\end{eqnarray}
where $S(t)$ denotes the step function, 
$v_n^{(1)}(t,\sigma)$, $u_n^{(1)}(t,\sigma)$, 
$v_n^{(2)}(t,\sigma,\epsilon)$ and $u_n^{(2)}(t,\sigma,\epsilon)$
admit power series expansions with respect to $\sigma$,
 and $E_{l}(t)$ is an exponentially small but 
highly oscillatory function of $t$ (more precisely, 
$E_{l}(t)={\rm O}(\epsilon)$ and is periodic with period $\sigma$). 

The difference equations for $v_n^{(l)}, u_n^{(l)}$ are obtained by 
substituting (\ref{2.2st}) into (1.1) and equating the terms with the 
same powers in 
$E_{l}(t)$. Particularly, $v_1^{(l)}, u_1^{(l)}$ obey a linear equation 
and their values at $t=T_l$ are necessary for fixing them. Such boundary 
values are provided with the aid of the Borel 
summable asymptotic expansions valid near a singularities in the
complex $t$-plane.
To see the method more in detail, let us focus on the time domains 
$-\infty <t<T_1$ and $T_1<t<T_2$.

\subsection{Melnikov perturbation}

The solution $v_0(t), u_0(t)$ describing the unstable manifold in 
the domain \break
$-\infty~<~t~<~T_1$ is given by the conventional Melnikov 
perturbation:
\begin{eqnarray}
v_0(t,\sigma) \equiv \sum_{i=0}^\infty \sigma^i v_{0i}(t) , \quad
u_0(t,\sigma) \equiv \sum_{i=0}^\infty \sigma^i u_{0i}(t) 
\nonumber
\end{eqnarray}
The lowest order solutions $u_{00}(t), v_{00}(t)$ are nothing but the 
solutions of the differential equation (\ref{eq:Harper0}) with the boundary 
condition 
(\ref{bc}). 
After fixing the origin of $t$ by 
$v_{0}(T)={\rm O}(\sigma^2)$, the terms
$(v_{00}(t),u_{00}(t))$ and $(v_{0i}(t),u_{0i}(t))$ $(i\ge 1)$
are found to 
have, respectively, logarithmic branch points and poles at 
\begin{eqnarray}
t=\frac{1}{\sqrt{k}} \left( n+{1\over 2}\right)\pi i,\ \ 
2T&+&\frac{1}{\sqrt{k}} \left( n+{1\over 2}\right)\pi i 
\nonumber\\
&& \mskip 50mu \quad (n=0,\pm 1, \pm 2, \cdots )
\end{eqnarray}
where $
T={1\over \sqrt{k}}\ln{1+\sqrt{k}\over \sqrt{1-k}}>0$.
On the other hand, by fixing the time origin as 
$v_{0}(0)=\pi/2+{\rm O}(\sigma^2)$,
the perturbative solution of the unstable manifold for $k=1$
have singularities at
\begin{eqnarray*}
t=\left( n+{1\over 2}\right)\pi i
\end{eqnarray*}
As one can easily seen from the definition of $T$, $T$ diverges as $k\to1-0$.
We choose the time origin so that the solution converges as $k\to1-0$
and the location of the singularities
in the limit of $k\to1-0$ matches with that for the case of $k=1$.
We remark that the choice of the time origin, $v_0(T)=0$
\footnote{
This choice gives the symmetric unstable manifold with respect to $u$ axis.
But it makes Stokes multiplier divergent and it indicates that 
the unstable manifold is not symmetric with respect to $u$ axis 
even within the perturabation approach.}, gives the 
solution which does not converge as $k\to1-0$ (see appendix~\ref{app i.c.}).

In the domain Re$ \ t<0$ of the complex time plane, $(v_0(t), u_0(t))$ is  
a reasonable approximation of the unstable manifold since 
individual terms of their expansions have no singularities there. 
The expressions valid in the domain ${\rm Re}~t >0$ can be 
constructed by
analytically continuing $v_0, u_0$ along any curve which does not cross 
the lines
$t=\pm is$ ($s>{\pi\over 2\sqrt{k}}$) emanating from the 
singularities of
$v_0, u_0$. Then, it is convenient to study 
the analytic 
continuation along the curve passing near the singular points since 
the divergence of the perturbative solutions is expected to magnify 
the small corrections.

\subsection{Behavior near a singularity}

Now we focus on the behavior of the perturbative solution near a singular 
point $t=t_1\equiv {i\pi \over 2\sqrt{k}}$.
The corrections $v_{0i}, u_{0i}$ have the Laurent expansions:
\begin{eqnarray}
\sigma^i
\begin{pmatrix}
v_{0i}(t)\cr u_{0i}(t)
\end{pmatrix}
={\sigma^i \over (t-t_1)^i}\sum_{l=0}^\infty
\begin{pmatrix}
a_l^{(i)}(t-t_1)^l \cr b_l^{(i)}(t-t_1)^l
\end{pmatrix}
\ , 
\quad
(i\ge 1, b_0^{(i)}\not= 0) \ .
\label{for appen}
\end{eqnarray}
Hence, higher order corrections are more singular. In such a case, the 
behavior of
the most divergent term is important. Because the most 
divergent contribution, e.g., to $v_0$ has the form:$\sum_i \sigma^i 
a_l^{(i)}/(t-t_1)^i$, 
it is convenient to introduce a new variable $z={t-t_1\over \sigma}$ and 
to rearrange the expansions as double expansions with respect to $\sigma$ 
and $z$.
Since one has
\begin{eqnarray}
&&\sigma^1 v_{01}(t,\sigma)= {a_0^{(1)}\over z}+\sigma {a_1^{(1)}\over z^0}+\cdots \cr
&&\sigma^2 v_{02}(t,\sigma)= {a_0^{(2)}\over z^2}+\sigma {a_1^{(2)}\over z^1}+\cdots \cr
&&\cdots\cdots\cdots\cdots\cdots\cdots\cdots\cdots\cdots\cdots\cdots \cr
&&\sigma^i v_{0i}(t,\sigma)= {a_0^{(i)}\over z^i}+\sigma {a_1^{(i)}\over z^{i-1}}+\cdots \nonumber
\end{eqnarray}
and the similar expression for $u_{0i}$, the perturbative solution can be 
reexpressed as
\begin{eqnarray}
\begin{pmatrix}
v_0(t,\sigma) \cr u_0(t,\sigma)
\end{pmatrix}
=\sum_{i=0}^\infty \sigma^i
\begin{pmatrix}
v_{0i}(t) \cr u_{0i}(t)
\end{pmatrix}
=
\begin{pmatrix}
v_{00}(t) \cr u_{00}(t)
\end{pmatrix}+
\sum_{l=0}^\infty \sigma^l
\begin{pmatrix}
V_{0l}(z) \cr U_{0l}(z)
\end{pmatrix}
\label{RewriteST}
\end{eqnarray}
where $V_{0l}, U_{0l}$ are given by
\begin{eqnarray}
V_{0l}(z)=\sum_{i=1}^{\infty}
\frac{\displaystyle a^{(i)}_l}{\displaystyle z^{i-l}}
\ , \qquad 
U_{0l}(z)=\sum_{i=1}^{\infty}\frac{\displaystyle b^{(i)}_l}
{\displaystyle z^{i-l}}
\label{eq:Borelst}
\end{eqnarray}
We remark that these series are divergent and are meaningful only as
asymptotic expansions. 

In short, the perturbative solution valid in the domain
$-\infty<{\rm Re} \ t<0$ turns out to be an asymptotic expansion near 
its singularity. Therefore, a careful treatment is necessary to analytically 
continue the solution across the line Re $t=0$.

\subsection{Borel resummation and analytic continuation}

Fortunately, the series (\ref{eq:Borelst}) are Borel summable. 
This can be shown as follows:
\vskip 5pt

\begin{enumerate}

\item[(i)] The equations for the Borel transforms ${\widetilde V}_{0l}(p), 
{\widetilde U}_{0l}(p)$ of $V_{0l}(z), U_{0l}(z)$ are derived from
the original equation (\ref{eq:Harper}) and the expansion (\ref{RewriteST}). 

\item[(ii)] The singularities of the Borel transforms are found with
the aid of the linearized inner equation. Indeed, 
according to the resurgence theory~\cite{Ecall,Gelfreich4}, 
the functional forms of the Borel transforms
${\widetilde V}_{0l}(p), {\widetilde U}_{0l}(p)$ 
are obtained from the linearized equation except some constants.

\item[(iii)] 
The Borel transformed equation is solved. This is carried
out as follows: The previous step determines the functional forms of 
${\widetilde V}_{0l}(p), {\widetilde U}_{0l}(p)$ up to a few arbitrary constants.
These constants are settled by comparing the power series expansions
of the guessed solutions with the power series solutions of the Borel 
transformed equation, which are numerically derived.

\end{enumerate}

\medskip

\noindent
Then, the Borel transforms ${\widetilde V}_{0l}(p), 
{\widetilde U}_{0l}(p)$ 
are found to have singularities along the lines $p=is, \
(s<-2\pi \ {\rm or} \ s>2\pi)$ and are analytic near the origin. 
The analyticity implies the existence of the asymptotic expansions 
of ${V}_{0l}(z), {U}_{0l}(z)$ at $|z|\to \infty$ in a certain sector. 
Hence, the divergent series (\ref{eq:Borelst}) are Borel summable. 

Now we consider the analytic continuations of ${V}_{0l}(z), {U}_{0l}(z)$.
Since negative powers of $z$ in $V_{0l}, U_{ol}$ are expressed as the inverse Borel transformation (namely the
Laplace transformation) of functions ${\widetilde V}_{0l}(p), 
{\widetilde U}_{0l}(p)$ which are analytic except on the lines
$p=is, \ 
(s<-2\pi \ {\rm or} \ s>2\pi)$, their analytic 
continuations from the sectors ${\rm Re}~ z <0,\ {\rm Im}~ z <0$ to 
${\rm Re} z >0,\ {\rm Im} z <0$, are obtained by appropriately changing the 
integration path of the Borel transformation.
Then, in the sector $0<{\rm Re} \ z,\ {\rm Im}~ z <0$, the solutions are expressed
as
\begin{eqnarray}
\begin{pmatrix}
v_{00}(t) \cr u_{00}(t)
\end{pmatrix}
+
\sum_{l=0}^\infty \sigma^l
\begin{pmatrix}
V_{0l}(z) \cr U_{0l}(z)
\end{pmatrix}
+
e^{-2\pi i z}
\sum_{l=0}^\infty \sigma^l
\begin{pmatrix}
V_{1l}(z) \cr U_{1l}(z)
\end{pmatrix}
+\cdots
\label{eqContinST}
\end{eqnarray}
where $V_{1l}(z), U_{1l}(z)$ are additional contributions and only the 
terms dominant for real $t$ are retained.  
More concretely, up to $\sigma^0$ and $z^{-2}$, one has
\begin{eqnarray}
&&e^{-2\pi i z}
\sum_{l=0}^\infty \sigma^l
\begin{pmatrix}
V_{1l}(z) \cr U_{1l}(z)
\end{pmatrix}
\cr
&&=
e^{-2\pi i z}\Biggl[
\Lambda_A
\left\{
\begin{pmatrix}
-{1\over z} \cr 
{1\over z}
\end{pmatrix}
-
\begin{pmatrix}
0 \cr 
\frac{1}{2z^{2}}
\end{pmatrix}
\right\} 
+
\Lambda_B
\left\{
\begin{pmatrix}
z
\cr
z
\end{pmatrix} 
+
\begin{pmatrix}
0
\cr
\frac{1}{2}
\end{pmatrix} 
+
\begin{pmatrix}
0
\cr
\frac{1}{12z}
\end{pmatrix} 
\right\}
+\cdots
\Biggr] 
\label{Match1}
\end{eqnarray}
where 
$$
\Lambda_\alpha=\Lambda_0^\alpha+
\sigma \Lambda_1^\alpha+\cdots
\quad
(\alpha=A, B)
$$
and the coefficients $\Lambda_i^\alpha$ are numerically
determined. Note that the analytic continuation (\ref{eqContinST}) simply
provides the asymptotic expansion near $t=t_1$ and that further 
investigation
is necessary to construct the analytically continued solution.

\subsection{Solutions in different domains}

Now we consider the solution for ${\rm Re} \ t >0$.
By comparing (\ref{2.2st}) with (\ref{Match1}) and reminding $\sigma z=t-t_1$,
one has
\begin{eqnarray}
E_1(t)={e^{-{\pi^2\over\sigma \sqrt{k}}}\over \sigma}e^{-{2\pi it \over \sigma}} \ ,
\end{eqnarray}
and, thus, $\epsilon=e^{-{\pi^2\over\sigma \sqrt{k}}}/\sigma$. 
Because $E_1(t)$ is exponentially small with respect to $\sigma$ and 
periodic in $t$ with period $\sigma$, equations for 
$(v_{n}^{(1)},u_{n}^{(1)})$ are obtained by substituting (\ref{2.2st}) 
to (\ref{eq:Harper}) and equating the terms with the same powers 
in $E_1(t)$. Substituting 
$v_{n}^{(1)}(t,\sigma)=\sum_{i=0}^\infty \sigma^iv_{ni}^{(1)}(t)$
and 
$u_{n}^{(1)}(t,\sigma)=\sum_{i=0}^\infty \sigma^iu_{ni}^{(1)}(t)$
into the so-obtained difference equations, one has a system of differential
equations $(n,i)\neq (0,0)$:
\begin{equation}
\begin{cases}
v_{ni}^{(1)\prime}(t)+
\displaystyle
{\sum_{j=2}^{i}\left(\frac{d}{dt}\right)^j\frac{v^{(1)}_{n\ i-j+1}}{j!}}
=-u_{ni}^{(1)}(t)\cos u_{00}(t)+f^{u}_{ni}(t)
\\
u_{ni}^{(1)\prime}(t)-
\displaystyle
{\sum_{j=2}^{i}(-1)^j\left(\frac{d}{dt}\right)^j\frac{u^{(1)}_{n\ i-j+1}}{j!}}
=kv_{ni}^{(1)}(t)\cos v_{00}(t)+f^{v}_{ni}(t)
\end{cases}
\label{sets}
\end{equation}
where $f^{v}_{ni}(t)$ and $f^{u}_{ni}(t)$ are functions of
$v_{jk}^{(1)}\ (j<n,\ k\le i)$, $v_{nk}^{(1)}\ (k<i)$,\ 
$v_{0k}\ (k\le i)$
and $u_{jk}^{(1)}\ (j<n,\ k\le i)$, $u_{nk}^{(1)}\ (k<i)$,\ 
$u_{0k}\ (k\le i)$, respectively.
The general solution of the system involves several constants of 
integration, which should be chosen so that the matching condition:
\begin{eqnarray}
\sum_{i=0}^\infty \sigma^i
\begin{pmatrix}
v_{1i}^{(1)}(t_1+\sigma z) \cr u_{1i}^{(1)}(t_1+\sigma z)
\end{pmatrix}
&=& \sigma
\sum_{l=0}^\infty \sigma^l
\begin{pmatrix}
V_{1l}(z) \cr U_{1l}(z)
\end{pmatrix}
\label{Match2}
\end{eqnarray}
is satisfied for small $\sigma$ and large $z$.
Another matching condition at $t=t_1^*$ similar to (\ref{Match2}) is
obtained by repeating the procedures explained in the previous subsections.
One can prove, from real analyticity of $(v_{0j},\ u_{0j})$, that 
the additional term obtained from the anlysis of $t=t_1^*$ is a conjugate 
to the term obtained from the analysis of $t=t_1$ 
(see appendix~\ref{app conjugate}).
These two matching conditions fix the constants of integration and, thus, 
the solutions $v_{n}^{(1)}(t,\sigma), u_{n}^{(1)}(t,\sigma)$.
As easily seen, the solutions have no singularity in $0 <{\rm Re} \
t <2T$.

This completes the derivation of the solution valid in the domain
$0 <{\rm Re} \ t <2T$.
	
When ${\rm Re}\ t$ exceeds $2T$, new terms will appear and 
the solutions valid in the domain ${\rm Re}\ t>2T$ can be obtained 
as follows:
\begin{itemize}
\item[(i)] The asymptotic expansions valid near $t=t_2\equiv 2T+i{\pi \over 2\sqrt{k}}$ are
derived as a function of $z'=(t-t_2)/\sigma$ and are analytically continued 
from the domain
${\rm Re} z' <0,\ {\rm Im} z' <0$ to 
${\rm Re} z' >0,\ {\rm Im} z' <0$ with the aid of the Borel resummation
and the resurgence theory.

\item[(ii)] The system of differential equations for the additional terms in ${\rm Re}\ t>2T$ 
is derived from (\ref{eq:Harper}) and (\ref{2.2st}). And its general solutions are calculated.

\item[(iii)] From the matching conditions such as (\ref{Match2}) 
near the singularities $t=t_2, t_2^*$, 
the constants of integration are fixed and one gets the desired solutions. 
\end{itemize}

\noindent
The final solution is shown to be analytic in the domain 
${\rm Re}\ t>2T$ and, thus, the description of 
the unstable manifold on the whole time domain is completed.
Note that, when $t$ passes the line Re$ \ t=2T$, the terms added in the 
domain $0<{\rm Re} \ t <2T$ generate new terms, which
are of order of $\epsilon^2$ and negligible for real $t$.
It will be proved in Sec.~6.

\subsection{Approximate unstable manifold}

Up to the order of $\sigma^3$ and $\sigma^1 \epsilon$, 
the unstable manifold is expressed as
\begin{eqnarray}
v_{u}(t)&&=
v_{00}(t)+\sigma^2 v_{02}(t) 
\nonumber \\
&&~+S(t)
{\rm Re}\left[
\frac{4\Lambda^{(1)}}{i\sigma}
e^{\frac{2\pi it_{1}}{\sigma}}
x_2(t)
e^{-\frac{2\pi it}{\sigma}}
\right]
\nonumber
\\
&&~+S(t-2T)
{\rm Re}\left[
\frac{4\Lambda^{(2)}}{i\sigma}
e^{\frac{2\pi it_{2}}{\sigma}}
\left\{
\left(\frac{T(k-1)^2-(1+k)}{4k(k-1)}\right)x_{1}(t)+x_2(t) 
\right\}
e^{-\frac{2\pi it}{\sigma}}
\right]
\nonumber
\\ \nonumber \\
u_{u}(t)
&&=
u_{00}(t)+\sigma \frac{y_{1}(t)}{2}+\sigma^{2}u_{02}(t)+\sigma^{3} 
\left(\frac{1}{2}u'_{02}(t)-\frac{1}{24}y''_{1}(t)\right)
\nonumber
\\
&&~+S(t)
{\rm Re}\biggl[
\frac{4\Lambda^{(1)}}{i\sigma}
e^{\frac{2\pi it_{1}}{\sigma}}
\left(y_{2}(t)+\sigma\frac{y'_{2}(t)}{2}\right)
e^{-\frac{2\pi it}{\sigma}}
\biggr]
\nonumber
\\
&&~+S(t-2T)
{\rm Re}\biggl[
\frac{4\Lambda^{(2)}}{i\sigma}
e^{\frac{2\pi it_{2}}{\sigma}}
\biggl\{ \left(\frac{T(k-1)^2-(1+k)}{4k(k-1)}\right)
\left(y_{1}(t)+\sigma\frac{y'_{1}(t)}{2}\right)
\nonumber \\
&& \mskip 375 mu +
\left(y_{2}(t)+\sigma\frac{y'_{2}(t)}{2}\right) \biggr\}
e^{-\frac{2\pi it}{\sigma}}
\biggr]
\nonumber
\end{eqnarray}
where $S(t)$ stands for the step 
function and the other functions are defined as follows:
\begin{eqnarray}
v_{00}(t)&=&-2\tan^{-1}\left[\sh{t}-\sqrt{k}\ch{t}
\right]
\nonumber
\\
u_{00}(t)&=&2\tan^{-1}\left[
\frac{\sqrt{k}}{-\sqrt{k}\sh{t}+\ch{t}}
\right] 
\label{sol:0th}
\\
v_{02}(t)&=&
-\frac{1}{24}\left[
x_{1}'(t)+x_{1}(t)\left\{kt-2\sqrt{k}
\frac{\left(\sinh\sqrt{k}t - \sqrt{k}\cosh\sqrt{k}t\right)
\left(\cosh\sqrt{k}t - \sqrt{k}\sinh\sqrt{k}t\right)}
{(1+k)\cosh^2 \sqrt{k}t-2\sqrt{k}\sinh\sqrt{k}t\cosh\sqrt{k}t}
\right\}
\right]
\nonumber
\\
u_{02}(t)&=&
\frac{1}{24}\left[
2y_{1}'(t)-y_{1}(t)\left\{kt-2\sqrt{k}
\frac{\left(\sinh\sqrt{k}t - \sqrt{k}\cosh\sqrt{k}t\right)
\left(\cosh\sqrt{k}t - \sqrt{k}\sinh\sqrt{k}t\right)}
{(1+k)\cosh^2 \sqrt{k}t-2\sqrt{k}\sinh\sqrt{k}t\cosh\sqrt{k}t}
\right\}
\right]
\nonumber\\
\label{sol:2nd}
\\
x_1(t)&=&-2\sqrt{k}
\frac{-\sqrt{k}\sinh\sqrt{k}t+\cosh\sqrt{k}t}
{1+\left(\sinh\sqrt{k}t-\sqrt{k}\cosh\sqrt{k}t\right)^2}
\nonumber
\\
y_1(t)&=&-2k
\frac{\sinh\sqrt{k}t-\sqrt{k}\cosh\sqrt{k}t}
{1+\left(\sinh\sqrt{k}t-\sqrt{k}\cosh\sqrt{k}t\right)^2}
\label{eq:x1}
\\
x_{2}(t)&=&-\frac{1}{4k}\left( \sh{t}-\sqrt{k}\ch{t} \right)
+\frac{1+k}{4\sqrt{k}} 
\frac{\ch{t}}
{1+\left(\sh{t}-\sqrt{k}\ch{t}\right)^2}
\nonumber
\\
&& \mskip 253mu+\frac{1-k}{8k}\left(t-\frac{i\pi}{2\sqrt{k}}\right) x_1(t)
\nonumber
\\
y_{2}(t)&=&\frac{1}{4\sqrt{k}}\left( \ch{t}-\sqrt{k}\sh{t} \right)
+\frac{1+k}{4\sqrt{k}} 
\frac{\ch{t}}
{1+\left(\sh{t}-\sqrt{k}\ch{t}\right)^2}
\nonumber
\\
&& \mskip 253mu+\frac{1-k}{8k}\left(t-\frac{i\pi}{2\sqrt{k}}\right) y_1(t)
\nonumber
\\
\label{eq:x2}
\end{eqnarray}
The coefficients $\Lambda^{(1)}$,\ $\Lambda^{(2)}$ are numerically
determined as
\begin{eqnarray}
\Lambda^{(1)}&=&\mskip 14 mu i4\pi^{3}A_{1}+
\sigma\pi^3\left\{
-(k-1)B_2-\frac{kt_1+1}{12}B_4
\right\}
\nonumber
\\
\Lambda^{(2)}&=&-i4\pi^{3}A_{1}+
\sigma\pi^3\left\{
-(k-1)B_2+\frac{kt_2-1}{12}B_4
\right\}
\end{eqnarray}
where $A_1=0.27893,\ B_2=0.14,\ B_4=3.503$.

\section{Melnikov Perturbation}\label{Sec3}

In this section, we start with the investigation of the perturbative 
solution, which provides a good approximation of the unstable manifold
for $t\to -\infty$. 
The lowest order terms in the expansion
$v_0(t,\sigma) \equiv \sum_{i=0}^\infty \sigma^i v_{0i}(t)$,
$u_0(t,\sigma) \equiv \sum_{i=0}^\infty \sigma^i u_{0i}(t)$
satisfy (\ref{eq:Harper0})
and the first and second order terms obey 
\begin{eqnarray}
&&v'_{01}(t)+u_{01}(t)\cos u_{00}(t)=-\frac{1}{2}v''_{00}(t)
\nonumber
\\
&&u'_{01}(t)- kv_{01}(t)\cos v_{00}(t)=\frac{1}{2}u''_{00}(t) 
\label{Meq:1st}
\\
&&v'_{02}(t)+u_{02}(t)\cos u_{00}(t) 
= -\frac{1}{2}v''_{01}(t)-\frac{1}{6}v'''_{00}(t)
+\frac{u^{2}_{01}(t)}{2}\sin u_{00}(t)
\nonumber
\\
&&u'_{02}(t) -k v_{02}(t) \cos v_{00}(t)
= \frac{1}{2}u''_{01}(t)-\frac{1}{6}u'''_{00}(t)
-k \frac{v_{01}^{2}(t)}{2}
\sin v_{00}(t)
\label{Meq:2nd}
\end{eqnarray}

\begin{figure}[t]
\label{fig:singular}
\begin{center}
\includegraphics[width=6cm,keepaspectratio,clip]{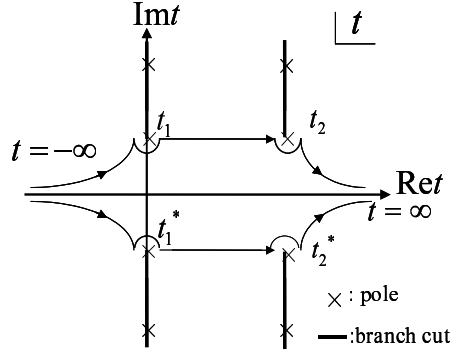}
\caption{Singular points (crosses) of the lowest order solution $v_{0}, \ u_{0}$ in the complex time domain.}
\end{center}
\end{figure}

The lowest order solution of the 
unstable manifold with $v_{00}(T)=0$ is given by (\ref{sol:0th}).
Because of the boundary conditions $v_{0n}(t) \to 0, 
\ u_{0n}(t) \to 0$ for $t\to -\infty$
and $v_{01}(T)=v_{03}(T)=0,\ v_{02}(T)=\frac{k^{3/2}}{12}\sqrt{1-k}$,
the first and the third order solutions are found to be
\begin{eqnarray*}
v_{01}(t)=0
\ , \qquad
u_{01}(t)=\frac{1}{2}y_{1}(t)
\end{eqnarray*}
\begin{eqnarray*}
v_{03}(t)=0
\ , \qquad
u_{03}(t)=\frac{1}{2}u_{02}'(t)-{1\over 24}y_1''(t)
\end{eqnarray*}
and the second order solution is given by (\ref{sol:2nd}).
It will be discussed in appendix~\ref{app i.c.}, we fix a time origin so that 
the final solution has no divergent term and the choice forbids a 
symmetric perturbative solution with respect to $u$ axis.
We observe that each order solution has two 
sequences of singular 
points in the complex $t$-plane (see Fig.~\ref{fig:singular}).
\begin{eqnarray}
t=\frac{1}{\sqrt{k}} \left( n+{1\over 2}\right)\pi i,\ \ 
2T&+&\frac{1}{\sqrt{k}} \left( n+{1\over 2}\right)\pi i 
\nonumber\\
&& \mskip 50mu \quad (n=0,\pm 1, \pm 2, \cdots )
\end{eqnarray}
where $
T={1\over \sqrt{k}}\ln{1+\sqrt{k}\over \sqrt{1-k}}>0$.
This implies that the perturbative solution approximates the unstable
manifold well in the domain: Re$ \ t<0$.

In the rest of this section, we examine the behavior of the perturbative 
solution near the singular point in the upper half plane:
$t_{1} \equiv {i\pi \over 2\sqrt{k}}$
closest to the real axis.
By substituting the Laurent expansion of each perturbative term to the
set of equations (\ref{sets}), we inductively get
\begin{eqnarray}
\sigma^i v_{0i}(t)=\frac{\sigma^i a^{(i)}_0}{(t-t_1)^i}+
\frac{\sigma^ia^{(i)}_1}{(t-t_1)^{i-1}}+\cdots,\ \ (i\ge 1)
\nonumber
\end{eqnarray}
and, thus,
\begin{eqnarray}
v_{0}(t,\sigma)-v_{00}(t)&=&\sum_{i=1}^\infty \sigma^i v_{0i}(t)
=\sum_{i=1}^{\infty}
\sig{l}\frac{a^{(i)}_{l}}{(t-t_{1})^{i-l}}\sigma^{i}
\label{div.exp}
\end{eqnarray}
This indicates that higher order terms with respect to $\sigma$ have
higher order poles at $t=t_{1}$ and that 
a careful investigation is necessary.
For this purpose, it is convenient to introduce 
a scaled variable
$z \equiv (t-{t_1})/\sigma$~\cite{Hakim,Tovbis,NakamuraHamada,Nakamurabook,NakamuraKushibe}.
Then, one finds that \break
$V_0(z,\sigma)\equiv
v_{0}(t_1+\sigma z,\sigma)-v_{00}(t_1+\sigma z)$
and 
$U_{0} (z,\sigma) \equiv 
u_{0}(t_1+\sigma z,\sigma)-u_{00}(t_1+\sigma z)$ 
can be 
expanded into power series with respect to $\sigma$:
\begin{eqnarray} 
&&\pmat{V_{0} (z,\sigma)\cr U_{0} (z,\sigma)}=
\sig{l} \sigma^{l} \pmat{V_{0l}(z)\cr U_{0l}(z)}\ 
\quad {\rm with} \
\pmat{V_{0l}(z)\cr U_{0l}(z)}=\sum_{i=1}^\infty {1\over z^{i-l}}
\pmat{a_l^{(i)}\cr b_l^{(i)}}
\label{AsympExpST}
\end{eqnarray}
and that 
$(V_{00}, U_{00})$, $(V_{01}, U_{01})$, $\dots$ correspond, respectively,
to the most divergent term, the next divergent term $\dots$ at $t=t_1$.
It turns out that the expansions (\ref{AsympExpST}) of $V_{0l}, U_{0l}$ 
are Borel summable asymptotic expansions which are valid in the 
sector Re$~z<0$. Hence, when one crosses the line Re$~z=0$,
new terms would be added. 
These aspects will be discussed in the next section.

\section{Inner Equation and Analytic Continuation of Solutions\label{secBorel}}

\subsection{Inner equation}\label{ssec5_1}

In terms of the variables
$V(z,\sigma)\equiv v(t_1+\sigma z,\sigma)-v_{00}(t_1+\sigma z)$ and 
\break
$U(z,\sigma)\equiv u(t_1+\sigma z,\sigma)-u_{00}(t_1+\sigma z)$,
the Harper map (\ref{eq:Harper}) reads as
\begin{eqnarray}
\Delta V (z,\sigma)
&=& -\sigma
\sin\left(U(z,\sigma)+ u_{00}(t_1+\sigma z)\right)
-\Delta v_{00}(t_1+\sigma z)
\nonumber
\\
\Delta U (z-1,\sigma)
&=& k\sigma
\sin\left(V(z,\sigma)+v_{00}(t_1+\sigma z)\right)
-\Delta u_{00}(t_1+\sigma (z-1))  \label{eq:inner}
\end{eqnarray}
where $\Delta$ stands for the difference operator: 
$\Delta f(z)\equiv f(z+1)-f(z)$.
This equation is used to study 
the asymptotic property of 
the solutions for 
$|z|\to\infty$~\footnote{The asymptotic behavior for $|z|\to\infty$
corresponds to that for $t\to t_1$ in the original varibale.}
and 
will be referred to as an inner equation.
On the other hand, the map (\ref{eq:Harper}) in the original time 
variable is referred to as an outer equation.

Let $V_{0}(z,\sigma),\ U_{0}(z,\sigma)$ be the solution of (\ref{eq:inner})
corresponding to the perturbative solution and let
\begin{equation}\label{InnerExpST}
\pmat{V_{0}(z,\sigma)\cr U_{0}(z,\sigma)}
=\sum_{l=0}^\infty \sigma^l \pmat{V_{0l}(z)\cr U_{0l}(z)}
\end{equation}
be their expansions with respect to $\sigma$, then the equations 
for $V_{0l},\ U_{0l}$
are obtained by substituting (\ref{InnerExpST}) into (\ref{eq:inner}).
The first two sets are
\begin{eqnarray}
\Delta V_{00}(z)&=&i\frac{e^{-iU_{00}(z)}}{z}-i\ln\left(1+\frac{1}{z}\right)
\nonumber
\\
\Delta U_{00}(z)&=&-i\frac{e^{i V_{00}(z+1)}}{z+1}+ i\ln\left(1+\frac{1}{z}\right)
\label{eq:inner0}
\\
\Delta V_{01}(z)
&=&\left[i\frac{k-1}{2}+\frac{U_{01}(z)}{z}\right] e^{-iU_{00}(z)}
-\frac{i}{2}(k-1)
\nonumber
\\
\Delta U_{01}(z)&=&
\left[\frac{V_{01}(z+1)}{z+1}-i\frac{1-k}{2}\right]e^{iV_{00}(z+1)}
+\frac{i}{2}(1-k)
\label{eq:inner1}
\end{eqnarray}
From these equations, one can directly obtain the expansion 
(\ref{AsympExpST}) discussed in the
previous section. 
Indeed, by posing the matching condition at $z=0$:
\begin{eqnarray}
{\rm Res} \left.\pmat{U_{0k}(z) \cr V_{0k}(z)}\right|_{z=0}
=
{\rm Res} \left.\pmat{v_{0\ \mskip -5mu k+1}(t) 
\cr u_{0\ \mskip -5mu k+1}(t)}\right|_{t=t_1}
\ ,
\end{eqnarray}
one has
\begin{eqnarray}
\begin{matrix}
&V_{00}(z)=& &~\displaystyle\frac{i}{12z^{2}}&
&\displaystyle-\frac{107i}{4320z^{4}}+{\rm O}\left(\frac{1}{z^5}\right)
\nonumber
\\
&U_{00}(z)=& \!\!\!\!\!\! \displaystyle-\frac{i}{2z} \!\!\!\!\!\!
&\displaystyle+\frac{i}{24z^{2}}&
\!\!\!\!\!
\displaystyle+\frac{i}{24z^{3}} \!\!\!\!\!\!
&\displaystyle- \frac{191i}{8640z^{4}} 
+{\rm O}\left(\frac{1}{z^5}\right)
\end{matrix}
\label{sol:inner00}
\\
\begin{matrix}
V_{01}(z)\!\!\! &=& &\displaystyle-\frac{i}{24}\frac{kt_{1}+1}{z}&
&\displaystyle+{\rm O}\left(\frac{1}{z^3}\right)\nonumber
\\
U_{01}(z)\!\!\!\!\!\! &=& 
\!\!\!\! \displaystyle\frac{i(k-1)}{4} \!\!\!\!\!\!
&\displaystyle+\frac{i}{24}\frac{k(t_{1}+1)}{z}& 
\!\!\!\!\!\! 
\displaystyle-\frac{i}{48}
\frac{k(t_1+1)}{z^{2}}
\!\!\!\!\!\!
&\displaystyle+{\rm O}\left(\frac{1}{z^3}\right)
\end{matrix}
\label{sol:inner01}
\end{eqnarray}
In the following subsections, we shall show that these expansions are Borel
summable and that their analytic continuations from the sector
${\rm Re}~ z <0,\ {\rm Im}~ z <0$ to 
${\rm Re}~ z >0,\ {\rm Im}~ z <0$ give necessary
information for the construction of the approximate
unstable manifold for $0<{\rm Re}~t < 2T$.
Note that , one can prove from the above asymptotic expansions and (\ref{eq:inner0}) that all the coefficient of 
$1/z$ exponents of $V_{00}(z),U_{00}(z)$ are purely imaginary.

\subsection{Borel transformation and analytic continuation}
\begin{figure}[t]
\begin{center}
\includegraphics[width=4cm,keepaspectratio,clip]{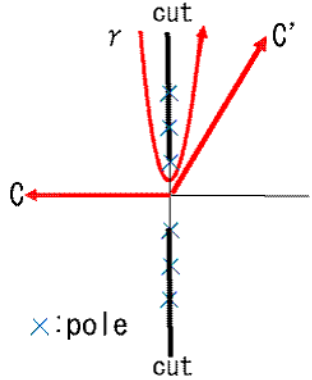}
\caption{Integration path in $p$-plane.}
\label{fig:int}
\end{center}
\end{figure}

We define their Borel transforms 
$\widetilde{V}_0(p),\ \widetilde{U}_0(p)$ by
\begin{eqnarray}
V_{00}(z)\equiv L[\widetilde{V}_0(p)](z) \equiv \lap \widetilde{V}_0(p)
\nonumber\\
U_{00}(z) \equiv L[\widetilde{U}_0(p)](z) \equiv \lap \widetilde{U}_0(p)
\label{borel trans}
\end{eqnarray}
The integration path is so chosen that the left hand sides are
regular for 
\break 
Re$~z\to -\infty$ since we are investigating the unstable 
manifold. 
As easily seen from (\ref{eq:inner0}), 
$\widetilde{V}_0(p), \ \widetilde{U}_0(p)$
satisfy 
\begin{eqnarray}
-i(e^{-p}-1) \widetilde{V}_{0}(p) &=&
1+ \int_{0}^{p} dp \left\{\sum_{n=1}^{\infty}
\frac{(- i)^{n}\widetilde{U}_{0}^{(*n)}(p)}{n!}\right\}
-\frac{1-e^{-p}}{p} 
\nonumber
\\
i(1-e^{p}) \widetilde{U}_{0}(p) &=&
1+ \int_{0}^{p}dp \left\{ \sum_{n=1}^{\infty}
\frac{i^{n}\widetilde{V}_{0}^{(*n)}(p)}{n!}\right\}
-\frac{e^{p}-1}{p} 
\label{innerborel}
\end{eqnarray}
where $\widetilde{V}_0^{(*n)}$ denotes the $n$th convolution 
recursively defined by
\begin{eqnarray}
\widetilde{V}_0^{(*n)}(p)=\int^{p}_{0} dx 
\widetilde{V}_{0}(p-x) \widetilde{V}_0^{(*(n-1))}(x)
\end{eqnarray}
The factors $(e^{-p}-1)$ and $(1-e^p)$ of (\ref{innerborel}) indicate 
that the Borel transforms 
$\widetilde{V}_{0}(p),\ \widetilde{U}_{0}(p)$ may have 
singularities at $p=\pm 2\pi n i$ ($n=1,2,\cdots$) and branch cuts 
starting from them (cf. Fig.~\ref{fig:int}).

Then, the analytic continuations of $V_{00}(z)$ and $U_{00}(z)$ from 
${\rm Re}~ z <0,\ {\rm Im}~ z <0$ to ${\rm Re}~ z >0,\ {\rm Im}~ z <0$ 
are obtained by rotating the integration path in the $p$-plane from C to C$'$
as shown in Fig.~\ref{fig:int}. 
Note that $V_{00}(z)$ and $U_{00}(z)$ should be analytically continued along a curve
passing below the singularity $z=0$ where they have no cut (cf. Fig.~\ref{fig:singular}).
This corresponds to 
counterclockwise rotation of $z$, e.g., along a semicircle below $z=0$ and, thus,
one should rotate the $p$-integration path in a clockwise way (cf. Fig.~\ref{fig:int}).
In short, the analytic continuation is given by
\begin{eqnarray}
V_{00}(z)=\int^{-\infty}_{0} dp \hspace*{1mm} e^{-pz} \hspace*{1mm}
\widetilde{V}_0(p)
\to
\int^{-\infty e^{i\theta}}_{0} dp 
\hspace*{1mm} e^{-pz} \hspace*{1mm}
\widetilde{V}_0(p)
-\int_{\gamma} dp \hspace*{1mm} e^{-pz} \hspace*{1mm}
\widetilde{V}_0(p)
\label{analytic}
\end{eqnarray}
where $\theta=\pi/2+\epsilon'$ with $\epsilon'$ being a small positive number.
The first term has the same asymptotic expansion with respect to $z$ as in the domain ${\rm Re}~ z <0,\ {\rm Im}~ z <0$
and the second term is the desired additional term. 

\subsection{Linearized equation}

If one can explicitly write down the Borel-transformed solution of
the inner equation, 
one has only to change the integration path.
But it is difficult to find the location and order of singularities
in a straightforward way. Instead, we investigate them with the aid
of the linearized inner equation.

As seen from the Borel transformed inner equation (\ref{innerborel}), 
singularities of the Borel transforms of $V_{00}(z)$ and $U_{00}(z)$ 
nearest to the real axis are located at $p=\pm 2\pi i$.
According to the resurgence theory~\cite{Ecall,Gelfreich4}, the order
of these singularities is given by the solution of the linearized
equation of (\ref{eq:inner0}):
\begin{eqnarray}
\Delta \Phi(z)&=& \frac{e^{- i U_{00}(z)}}{z}\Psi(z)
\ , \qquad
\Delta \Psi(z)= \frac{e^{iV_{00}(z+1)}}{z+1}\Phi(z+1) \ .
\label{inner10b}
\end{eqnarray}
More precisely, let $(V_A(z),U_A(z))$ and $(V_B(z),U_B(z))$ be 
two linearly independent solutions of (\ref{inner10b}) and let 
$({\widetilde V}_A(p),
{\hat U}_A(p))$ and $({\hat V}_B(p),{\hat U}_B(p))$
be the functions satisfying
\begin{equation}
e^{-2\pi iz}
\pmat{V_\alpha(z)\cr U_\alpha(z)}=-\int_\gamma dp e^{-pz}
\pmat{{\hat V}_\alpha(p)\cr {\hat U}_\alpha(p)}
\qquad (\alpha=A,B),
\label{LinearBorelST}
\end{equation}
then, with appropriately chosen constants $\Lambda_0^A$ and
$\Lambda_0^B$, the Borel transforms of 
$V_{00}(z)$ and $U_{00}(z)$ are given by
\begin{equation}
\pmat{{\widetilde V}_0(p)\cr {\widetilde U}_0(p)}
=
\Lambda_0^A
\pmat{{\hat V}_A(p)\cr {\hat U}_A(p)}
+\Lambda_0^B
\pmat{{\hat V}_B(p)\cr {\hat U}_B(p)}
+\pmat{R_v(p)\cr R_u(p)}
\label{LinearMatchST}
\end{equation}
in a neighborhood of a singular point $p=2\pi i$, where the functions $R_v(p)\ , R_u(p)$
are regular near $p=2\pi i$.

With the aid of this observation, the Borel transforms 
${\widetilde V}_0(p)$ and 
${\widetilde U}_0(p)$ are determined as follows: At first, with the 
aid of the 
formulas\footnote{We have chosen the branch cut of the logarithm along 
the positive imaginary axis as shown
in Fig.~\ref{fig:int}. Note that this choice fixes the Stokes line on 
the negative imaginary axis in the 
$z$-plane, which corresponds to the line in the original $t$-plane 
joining $t_1$ and its complex conjugate.
}
\begin{eqnarray}
&&-\int_{\gamma} dp \hspace*{1mm} e^{-pz} \hspace*{1mm}
\left[\frac{(j-1)!}{(p-2\pi ni)^{j}}\right]
=
-2\pi i(-z)^{j-1}e^{-2\pi inz}
\nonumber
\\
&&-\int_{\gamma} dp \hspace*{1mm} e^{-pz} \hspace*{1mm}
\left[\frac{(p-2\pi in)^{j}\ln(p-2\pi in)}{j!}\right]
=
\frac{-2\pi i}{z^{j+1}}e^{-2\pi inz}
\label{trivialST}
\end{eqnarray}
and the relation (\ref{LinearBorelST}), one derives approximate functional
forms of ${\hat V}_\alpha(p)$, ${\hat U}_\alpha(p)$ 
($\alpha=A,B$) near $p=2\pi i$ from the asymptotic $1/z$-expansion of the
solution of (\ref{inner10b}).

As pointed out at the end of Sec.~\ref{ssec5_1}, 
all the coefficients of $1/z$-expansions of $V_{00}(z)$ and $U_{00}(z)$ are
purely imaginary and this implies that ${\widetilde V}_0(p)$, 
${\widetilde U}_0(p)$ are purely imaginary for real $p$.
By taking this property into account, the functional forms of
${\widetilde V}_0(p)$, ${\widetilde U}_0(p)$
are guessed from (\ref{LinearMatchST}).
Next, the power series solution of the Borel transformed inner 
equation (\ref{innerborel}) is obtained numerically. 
Finally, the coefficients $\Lambda_0^A$ and $\Lambda_0^B$ are determined by 
comparing the power series solution of ${\widetilde V}_0(p)$, 
${\widetilde U}_0(p)$ 
and the power series expansion of their guessed functional forms.
This calculation will be carried out in the next subsection.

In the rest of this subsection, we give a plausible argument why the Borel transformed inner solution 
near the singularity can be
determined from the linearized equation. We restrict ourselves to the
case where the Borel
transformed solution has the form\footnote{For the H\' enon map, 
Gelfreich and Sauzin~\cite{Gelfreich4} have shown that the solution 
of the Borel transformed inner equation has such a form in a neighborhood of the 
singularity.}:
\begin{equation}
\pmat{{\widetilde V}_0(p)\cr {\widetilde U}_0(p)}
=
\sum_{n=1}^\infty
\pmat{g_n^v(p-2\pi ni)+h_n^v(p-2\pi ni)\ln(p-2\pi ni) \cr
g_n^u(p-2\pi ni)+h_n^u(p-2\pi ni)\ln(p-2\pi ni)}
\end{equation}
where $g_n^\alpha(p)$ ($\alpha=u,v$) are meromorphic with a pole only at the origin
and $h_n^\alpha(p)$ ($\alpha=u,v$) are analytic and 
$e^{-pz}h_n^\alpha(p) ({\rm Im}~ z<0)$ are rapidly decreasing for $p\to i\infty$.
Then, their contour integrals along the path $\gamma$ are given by
\begin{equation}
-\int_\gamma dp e^{-pz}
\pmat{{\widetilde V}_0(p)\cr {\widetilde U}_0(p)}
=
\sum_{n=1}^\infty e^{-2\pi i nz}
\pmat{V_{n0}(z)\cr U_{n0}(z)}
\label{5-15}
\end{equation}
where
\begin{eqnarray}
\pmat{V_{n0}(z)\cr U_{n0}(z)}={2\pi\over i}
\left[e^{2\pi i nz}
{\rm Res}\left[e^{-pz}
\pmat{g_n^v(p)\cr g_n^u(p)}\right]_{p=0}
+\int_0^\infty
dq e^{-i qz}\pmat{h_n^v(iq)\cr h_n^u(iq)}
\right]
\ \label{hol}.
\end{eqnarray}
The first terms of the right hand side of (\ref{hol}) are polynomials of $z$ 
and the second terms can be
expanded into a power series with respect to $1/z$.
On the other hand, since
$$
\pmat{V_{00}(z)-\int_\gamma dp e^{-pz}{\widetilde V}_0(p)
\cr U_{00}(z)-\int_\gamma dp e^{-pz}{\widetilde U}_0(p)}
=
\pmat{V_{00}(z)\cr U_{00}(z)}+
\sum_{n=1}^\infty e^{-2\pi i nz}
\pmat{V_{n0}(z)\cr U_{n0}(z)}
$$
and $(V_{00}(z), U_{00}(z))$ are solutions of (\ref{eq:inner0}), 
by substituting the above expansion into (\ref{eq:inner0})
and comparing the terms with the same powers in $e^{-2\pi i z}$,
one finds that $V_{10}(z)$, $U_{10}(z)$ satisfy the linearized equation
(\ref{inner10b}) and, thus, are linear combinations of 
$(V_A(z), U_A(z))$ and $(V_B(z), U_B(z))$.
As a consequence, there exist constants $\Lambda_0^A$ and $\Lambda_0^B$
such that 
\begin{eqnarray}
&&-\int_\gamma dp e^{-pz}
\pmat{g_1^v(p-2\pi i)+h_1^v(p-2\pi i)\ln(p-2\pi i) \cr
g_1^u(p-2\pi i)+h_1^u(p-2\pi i)\ln(p-2\pi i)}
\cr
&&~~=e^{-2\pi i z}
\pmat{V_{10}(z)\cr U_{10}(z)}=
e^{-2\pi i z}
\left\{\Lambda_0^A
\pmat{V_A(z)\cr U_A(z)}+
\Lambda_0^B
\pmat{V_B(z)\cr U_B(z)}
\right\}
\ .
\end{eqnarray}
This relation corresponds to (\ref{LinearMatchST}) and 
implies that the functional form of the Borel transformed
solution ${\widetilde V}_0(p), \ {\widetilde U}_0(p)$ can be
determined by the linearized equation.

\subsection{Determination of ${\widetilde V}_{0}(p),
\ {\widetilde U}_{0}(p)$ }

Now we calculate $\widetilde{V}_{0}(p),\ \widetilde{U}_{0}(p)$
following the procedure described in the previous subsection.
Firstly, the linearized equation (\ref{LinearBorelST}) has
two linearly independent solutions $(V_A(z), U_A(z))$ and
$(V_B(z), U_B(z))$, which can be expanded for large $z$ as 
\begin{eqnarray}
&V_A(z)=\displaystyle-\frac{1}{z}+\frac{1}{6z^{3}} +\cdots 
\ , \mskip 60mu
&U_A(z)=\displaystyle 
\frac{1}{z}-\frac{1}{2z^{2}}+\frac{1}{12z^{3}}+\cdots 
\cr
&V_B(z)=\displaystyle 
z-\frac{1}{540z^{3}}+\cdots 
\ , \mskip 10mu
&U_B(z)=\displaystyle 
z+\frac{1}{2}+\frac{1}{12z}-\frac{1}{24z^{2}}+\frac{11}{2160z^{3}}+\cdots
\nonumber\\
\end{eqnarray}
Then, (\ref{trivialST}) leads to
\begin{eqnarray}
{\hat V}_A(p)&=&{1\over 2\pi i}\ln(p-2\pi i)
+\cdots
\cr
{\hat V}_B(p)&=&{1\over 2\pi i}
{1\over (p-2\pi i)^2}
+\cdots
\end{eqnarray}
where the omitted terms correspond to $z^{-m}$ ($m\ge 2$).
Hence, one has
\begin{eqnarray}
{\widetilde V}_0(p)&=&
\Lambda_0^A
{\hat V}_A(p)
+\Lambda_0^B
{\hat V}_B(p)+\cdots
\cr
&=&{1\over 2\pi i}\left\{
{\Lambda_0^B \over (p-2\pi i)^2}
+
\Lambda_0^A
\ln(p-2\pi i)\right\}
+\cdots
\end{eqnarray}
where the rest part is either regular or less singular at $p=2\pi i$.
As mentioned at the end of Sec.~\ref{ssec5_1}, all the coefficients
of the $1/z$-expansion of $V_{00}(z)$ are purely imaginary and, thus,
${\widetilde V}_{0}(p)$ is purely imaginary for real $p$.
This property determines a part of the regular components of
${\widetilde V}_0(p)$ and, as a result, one can guess
\begin{eqnarray}
{\widetilde V}_0(p)&=&
{1\over 2\pi i}\Biggl\{
{\Lambda_0^B \over (p-2\pi i)^2}
+{{\Lambda_0^B}^* \over (p+2\pi i)^2}
+
\Lambda_0^A\ln(p-2\pi i)
+{\Lambda_0^A}^*
\ln(p+2\pi i)\Biggr\}
+\cdots
\cr
&=& {\rm Re}\Lambda_0^B \ f_1^{(R)}(p)+
{\rm Im}\Lambda_0^B \ f_1^{(I)}(p)
-{\rm Re}\Lambda_0^A f_{-1}^{(R)}(p)
-{\rm Im}\Lambda_0^A f_{-1}^{(I)}(p)+\cdots
\label{GuessST}
\end{eqnarray}
where the following auxiliary functions are introduced
\begin{eqnarray}
f^{(R)}_{j}(p)&=&\frac{j!}{2\pi i}
\left(\frac{1}{(p-2\pi i)^{j+1}}+\frac{1}{(p+2\pi i)^{j+1}}\right)
\quad (j=0,1,2,\cdots)
\cr
f^{(I)}_{j}(p)&=&\frac{j!}{2\pi}
\left(\frac{1}{(p-2\pi i)^{j+1}}-\frac{1}{(p+2\pi i)^{j+1}}\right)
\quad (j=0,1,2,\cdots)
\cr
f^{(R)}_{-1}(p)
&=&
\frac{i}{2\pi}
\left[\ln(p-2\pi i) + \ln(p+2\pi i)\right]
\cr
f^{(I)}_{-1}(p)
&=&
-\frac{1}{2\pi}
\left[\ln(p-2\pi i) - \ln(p+2\pi i)\right]
\nonumber 
\end{eqnarray}
We note that the above auxiliary functions have power series of $p$ with purely imaginary coefficients.
One can easily show that the neglected terms of (\ref{GuessST}) are 
either regular or less singular at $p=\pm 2\pi i$.
As easily seen, (\ref{GuessST}) admits the power series expansion near $p=0$:
\begin{eqnarray}
{\widetilde V}_0(p)&=& M+\sum_{n=0}^\infty\left\{
{i{\rm Im}\Lambda_0^B \over 4\pi^3}(2n+2)+{i{\rm Im}
\Lambda_0^A\over \pi(2n+1)}
\right\}
(-1)^n \left({p\over 2\pi}\right)^{2n+1}
\cr
&&+\sum_{n=1}^\infty\left\{
{i{\rm Re}\Lambda_0^B \over 4\pi^3}(2n+1)+{i{\rm Re}
\Lambda_0^A \over 2\pi n}
\right\}
(-1)^n \left({p\over 2\pi}\right)^{2n} \ ,
\label{PowerST1}
\end{eqnarray}
where $M=-\frac{i{\rm Im}\Lambda^A_0}{4}+\frac{i{\rm Re}\Lambda^B_0}{4\pi^3}
-\frac{i{\rm Re}\Lambda^A_0}{2\pi}\log (4\pi^2)$.
In a similar way, one can guess
\begin{eqnarray}
{\widetilde U}_0(p)&=& M'+\sum_{n=0}^\infty\left\{
{i{\rm Im}\Lambda_0^B \over 4\pi^3}(2n+2)-{i{\rm Im}
\left({\Lambda_0^A+\Lambda_0^B/12}\right) \over \pi(2n+1)}
+{i{\rm Re}\Lambda_0^B \over 4\pi^2}
\right\}
(-1)^n \left({p\over 2\pi}\right)^{2n+1}
\cr
&&+\sum_{n=1}^\infty\left\{
{i{\rm Re}\Lambda_0^B \over 4\pi^3}(2n+1)-{i{\rm Re}
\left({\Lambda_0^A+\Lambda_0^B/12}\right) \over 2\pi n}
-
{i {\rm Im}\Lambda_0^B\over 4\pi^2}
\right\}
(-1)^n \left({p\over 2\pi}\right)^{2n} \ ,
\label{PowerST2}
\end{eqnarray}
where $M'=M+\frac{i{\rm Im}~(2\Lambda^A_0+\Lambda^B_0/12)}{4}
-\frac{i{\rm Im}\Lambda^B_0}{4\pi^2}$.

Now we numerically solve the Borel transformed inner equation
(\ref{innerborel}).
By substituting the power series expansions:
\begin{eqnarray}
\widetilde{V}_{0}(p)&\equiv&\sig{n}a_{n}p^{n}
, \quad
\widetilde{U}_{0}(p)\equiv\sig{n}b_{n}p^{n}
; \quad a_{0}=0 , \quad b_{0}=- \frac{i}{2}
\nonumber
\end{eqnarray}
into (\ref{innerborel}) and comparing term by term, the coefficients 
$a_{n},\ b_{n}$ are determined recursively and one has
\begin{eqnarray}
ia_{2n+1}(-1)^{n}(2\pi)^{2n+1} &\rightarrow& -(2n+2)A_1
+\frac{A_2}{2n+1}
\quad (n\to \infty)
\nonumber
\\
ia_{2n}&=&0,\ \ \ \ (\forall n)
\label{numericalST1}
\\
ib_{2n+1}(-1)^{n}(2\pi)^{2n+1} &\rightarrow& -(2n+2)A_1
+\frac{A_2}{2n+1}
\quad (n\to \infty)
\nonumber
\\
ib_{2n}(-1)^{n}(2\pi)^{2n} &\rightarrow& A_3
\quad (n\to \infty)
\label{numerical1}
\end{eqnarray}
where $A_1=0.27893, \ A_2=0.417, \ A_3=0.87628$.

By comparing (\ref{PowerST1}) with (\ref{numericalST1}), 
and (\ref{PowerST2}) with (\ref{numerical1}), one finds that
$\Lambda_0^A$ and $\Lambda_0^B$ are purely imaginary 
and
\begin{eqnarray}
\Lambda_0^B= 4\pi^3 i A_1 \ , \quad
\Lambda_0^A+{\Lambda_0^B\over 12}=-\Lambda_0^A=\pi i A_2 \ , \quad
\Lambda_0^B= 4\pi^2 i A_3 \ . 
\label{NumericalRelation}
\end{eqnarray}
Because of the first and third equalities of (\ref{NumericalRelation}), the ratio $A_3/A_1$ has to be
$\pi \Lambda_0^B/\Lambda_0^B=\pi$ and the evaluated
values give an excellent agreement: $A_3/A_1= 3.1416$. Also,
the first and second equalities require $(6A_2)/(\pi^2A_1)$ to be unity
and the evaluated values are consistent:$(6A_2)/(\pi^2A_1)=0.909$\footnote{
The numerical evaluations of $A_1,\ A_3$ are quite robust.
On the other hand, $A_2$ is sensitive to an error of $A_1$ 
since the coefficients of $1/(2n+1)$ are fitted after subtracting 
the leading terms of order $n$.
However, we observe that the coefficients of $1/(2n+1)$ in 
$a_{2n+1}(-1)^{n}(2\pi)^{2n+1}$ and
$b_{2n+1}(-1)^{n}(2\pi)^{2n+1}$ 
are always the same, or the second equation of (\ref{NumericalRelation}) 
holds. And the value of $A_2$ are fitted so that the first and second 
equations are well satisfied. 
}.

Here we remark about the contributions from a singularity at $p=ip_c$
($p_c>2\pi$). If the singularity is of the same type as that at $p=2\pi i$,
the contributions to the coefficient of $p^n$ are exponentially smaller 
by a factor of $\left({2\pi \over p_c}\right)^n$ than those from $p=2\pi i$.
Thus, such contributions do not affect the present analysis and, at the
same time, are difficult to be evaluated.

In short, following the procedure discussed in the previous subsection, 
we have obtained the Borel transforms $\widetilde{V}_0(p), \ \widetilde{U}_0(p)$:
\begin{eqnarray}
\widetilde{V}_{0}(p)&=& -{\Lambda_0^A\over i}f^{(I)}_{-1}(p)
+{\Lambda_0^B\over i}f_{1}^{(I)}(p)
\nonumber
\\
\widetilde{U}_{0}(p)&=& -{\Lambda_0^A\over i}f^{(I)}_{-1}(p)+
{\Lambda_0^B\over i}\left(f_{1}^{(I)}(p)-{1\over 2}f_{0}^{(I)}(p)
\right)
\label{sol:numerical0}
\end{eqnarray}

\subsection{Borel transforms of $V_{01}(z)$ and $U_{01}(z)$}

Let $V_{01}^-(z),\ U_{01}^-(z)$ be the sum of negative powers of $z$ 
in the asymptotic expansion $V_{01}(z),\ U_{01}(z)$, and
let $\widetilde{V}_1(p), \widetilde{U}_1(p)$
be the Borel transforms of $V^-_{01}(z),\ U^-_{01}(z)$ 
in the sector including $z=-\infty$, 
then they provide the following additional terms in the other sector
including $z=+\infty$
\begin{eqnarray}
-\int_{\gamma} dp \hspace*{1mm} e^{-pz} \hspace*{1mm}
\pmat{\widetilde{V}_{1}(p)\cr \widetilde{U}_{1}(p) }
=e^{-2\pi iz} 
\pmat{V_{11}(z)  \cr U_{11}(z) }
+\cdots
\label{borel11ST}
\end{eqnarray}
where the neglected terms are proportional to a factor of 
$e^{-i p_c z}$ ($p_c>2\pi$).

Similarly to the construction of $\widetilde{V}_0(p), \ \widetilde{U}_0(p)$,
one can get  $\widetilde{V}_1(p), \ \widetilde{U}_1(p)$ 
from $V_{11}(z), \ U_{11}(z)$, which obey:
\begin{eqnarray}
\Delta V_{11}(z)&=&
\frac{U_{11}(z) +\frac{k-1}{2}z U_{10}(z)}{z}e^{- iU_{00}(z)}
-i\frac{U_{10}(z) U_{01}(z)}{z}e^{- i U_{00}(z)}
\nonumber
\\
\Delta U_{11}(z-1)&=&
\frac{V_{11}(z)+\frac{1-k}{2}z V_{10}(z)}{z}e^{iV_{00}(z)}
+ i\frac{V_{10}(z) V_{01}(z)}{z}e^{i V_{00}(z)}
\label{eq:11}
\end{eqnarray}
where $V_{10}, U_{10}$ are introduced in (\ref{5-15}).

Because of the linearity of (\ref{eq:11}), $V_{11}$ and $U_{11}$ are given by
\begin{eqnarray}
\pmat{V_{11}(z)\cr U_{11}(z)}= \Lambda_0^A \pmat{V_{11}^A(z)\cr U_{11}^A(z)}
+\Lambda_0^B \pmat{V_{11}^B(z)\cr U_{11}^B(z)}
+\Lambda_1^A \pmat{V_A(z)\cr U_A(z)}
+\Lambda_1^B \pmat{V_B(z)\cr U_B(z)}
\ \ \ 
\end{eqnarray}
where the constants $\Lambda_0^\alpha$ ($\alpha=A,B$) are introduced in 
the previous subsection, $\Lambda_1^\alpha$ ($\alpha=A,B$) are new
constants, and $V_{11}^\alpha(z), U_{11}^\alpha(z)$ ($\alpha=A,B$) are 
solutions of 
\begin{eqnarray}
\Delta V_{11}^\alpha(z)&=&
\frac{U_{11}^\alpha(z) +\frac{k-1}{2}z U_\alpha(z)}{z}e^{- iU_{00}(z)}
-i\frac{U_\alpha(z) U_{01}(z)}{z}e^{- i U_{00}(z)}
\nonumber
\\
\Delta U_{11}^\alpha(z-1)&=&
\frac{V_{11}^\alpha(z)+\frac{1-k}{2}z V_\alpha(z)}{z}e^{iV_{00}(z)}
+ i\frac{V_\alpha(z) V_{01}(z)}{z}e^{i V_{00}(z)}
\label{eq:11ST}
\end{eqnarray}
such that the asymptotic expansions of $V_{11}^\alpha(z)$ ($\alpha=A,B$) have no terms proportional
to $z$ nor $1/z$. Then, asymptotic expansions 
of $V_{11}(z), \ U_{11}(z)$ are obtained by 
exactly the same procedure as in the previous subsection and, for large $|z|$, 
we have
\begin{eqnarray}
\pmat{V_{11}(z)\cr U_{11}(z)}\approx
\pmat{
\Lambda^B_{1}z + \frac{k-1}{6}\Lambda^B_{0}z^{2}
\cr
\Lambda^B_{1}z - \frac{k-1}{6}\Lambda^B_{0}(z^{2}+z)
}
\label{11estimationST}
\end{eqnarray}
where the terms of order $z^{-m}$ ($m\ge 0$) are omitted since the
numerical estimation of their contributions to ${\widetilde V}_1(p),
{\widetilde U}_1(p)$ is very difficult\footnote{Such an estimation 
requires a separation of the terms of order $1/n$ from the coefficients
$a_n$, $b_n$ of the power series expansion of ${\widetilde V}_1,
{\widetilde U}_1$. But, since their leading order terms are
proportional to  $n^2$, it is very difficult.}.

Now we go back to the equation (\ref{eq:inner1}) of $V_{01}(z), U_{01}(z)$.
Because it is linear, one gets (See appendix~\ref{app01} for more details.)
\begin{eqnarray}
\pmat{V^-_{01}(z) \cr U^-_{01}(z)}=
i\left\{
\frac{kt_1 +1}{24}\pmat{V_{A}(z) \cr U_{A}(z)}+
\frac{k-1}{2}\pmat{V_{B}(z)-z \cr U_{B}(z)-(z-\frac{1}{2})}
\right\}.
\label{01appe}
\end{eqnarray}
As one can see from the definition of 
$\pmat{V_{\alpha}(z) \cr U_{\alpha}(z)}$,
the Borel transform of it has real values for real $p$.
On the other hand, (\ref{11estimationST}) shows that 
$\widetilde{V}_1(p), \ \widetilde{U}_1(p)$ have the first and second order 
poles at $p=2\pi i$. These observations imply
\begin{eqnarray}
\pmat{\widetilde{V}_{1}(p) \cr \widetilde{U}_{1}(p)}=
-4\pi^4 (k-1) B_1 f_2^{(I)}&&\pmat{1 \cr -1}
-\left(
\frac{kt_1+1}{12}\pi^3 B_4 +(k-1)\pi^3 B_2
\right)f_1^{(R)}(p)
\pmat{1 \cr 1}
\nonumber
\\
&&-\pi^3 (k-1)B_3f_1^{(I)}(p) \pmat{0 \cr 1}
\label{sol:numerical1ST}
\end{eqnarray}

\begin{eqnarray}
\pmat{V_{11}(z) \cr U_{11}(z)}&&=
4\pi^4 i(k-1) B_1 z^2\pmat{1 \cr -1}
\nonumber
\\
&&-\left(
\frac{kt_1+1}{12}\pi^3 B_4 +(k-1)\pi^3 B_2
\right)
z\pmat{1 \cr 1}
-\pi^3 i (k-1) B_3 z\pmat{0 \cr 1}
\label{sol:nume1}
\end{eqnarray}

The constants $B_1$, $B_2$, $B_3$ and $B_4$ can be numerically 
evaluated by considering the Borel transformations of 
$V_A,\ U_A,\ V_B-z$ and $U_B-z-1/2$
(for more detail, see Appendix~\ref{app01}).
Then, we have
\begin{eqnarray}
B_1=0.01480,\  \ B_2=0.14,\ B_3=0.186,\ B_4=3.503
\label{Bs}
\end{eqnarray}
By substituting (\ref{sol:nume1}) into (\ref{11estimationST}), one gets
\begin{eqnarray}
\Lambda^B_{0}&=&24\pi^{4}iB_{1}=6\pi^{3}iB_{3}
\nonumber
\\
\Lambda^B_{1}&=&-(k-1)\pi^3 B_2-\frac{kt_1 +1}{12}\pi^3 B_4.
\label{StokesV}
\end{eqnarray}
Eq.~(\ref{StokesV}) and $\Lambda^B_{0}=i4\pi^{3}A_{1}$ imply 
two relations among $A_1$, $B_2$ and $B_4$, which are satisfied by 
the present numerical estimations rather well:
\begin{eqnarray}
{A_1\over 6 \pi B_1}=0.99987\simeq 1 \qquad
{2A_1\over 3 \pi B_3}=0.99974\simeq 1 \nonumber \ .
\end{eqnarray}
Hence, we obtain
\begin{eqnarray}
\widetilde{V}_{1}(p)&=&
i{\Lambda_0^B (k-1)\over 6}f_{2}^{(I)}(p)+
\Lambda_1^B f_{1}^{(R)}(p)
\nonumber
\\
\widetilde{U}_{1}(p)&=&-i{\Lambda_0^B (k-1)\over 6}
\left(f_{2}^{(I)}(p)-f_{1}^{(I)}(p)\right)
+\Lambda_1^B f_{1}^{(R)}(p)
\label{sol:borel11ST}
\end{eqnarray}

\subsection{Asymptotic expansions of the additional terms}

In summary, we have shown, up to the order of $\sigma^1$,  that
$V(z,\sigma), U(z,\sigma)$ acquire new terms in the sector 
including $z=+\infty$ 
\begin{eqnarray}
\pmat{V(z,\sigma)\cr U(z,\sigma)}
=\pmat{V_{00}(z)+\sigma V_{01}(z) 
\cr U_{00}(z)+\sigma U_{01}(z)}
+e^{-2\pi iz}
\pmat{V_{10}(z)+\sigma V_{11}(z)
\cr U_{10}(z)+\sigma U_{11}(z)}
+\cdots \ \ \ \ \ \ 
\end{eqnarray}
where the first term corresponds to the perturbative solution studied
in the previous section and the second term is
the additional term:
\begin{eqnarray}
V_{10}(z)+\sigma V_{11}(z)&\approx&
\left(-\frac{\Lambda_0^A}{z}+\Lambda_0^B z \right)
+\sigma\left(\Lambda^B_{1}z + \Lambda^B_{0} \frac{k-1}{6} z^{2}\right)
\label{5-38}
\\
U_{10}(z)+\sigma U_{11}(z) &\approx& \left[
\frac{\Lambda_0^A}{z}
+\Lambda_0^B\left(z+\frac{1}{2}+\frac{1}{12z}\right)
\right]
+\sigma \left[
\Lambda^B_{1}z - \Lambda^B_{0}\frac{k-1}{6}(z^{2}+z)
\right]
\nonumber \\
\label{5-39}
\end{eqnarray}
Although the functional forms of $V_{10},U_{10},V_{11},U_{11}$ can be derived
from their linear equations, the resurgence theory and Borel resummation 
play an essential role for the determination of the coefficients 
$\Lambda^A_0, \Lambda^B_0, \Lambda^B_1$. 

In the next section, only the terms of order of $z$ and $z^2$ are used 
to derive the solutions of the outer equation in the domain 
Re~$t>0$ since $\sigma/z$-terms are not determined 
because of the nemerical difficulty.
However, it should be noted that the existence of $1/z$ terms does show the fact that the 
point $p=2\pi i$ is a branch point of the Borel transforms of $V(z,\sigma),U(z,\sigma)$.
Before closing this section, we remark that (\ref{5-38}) can be
rewritten as
\begin{eqnarray}
V_{10}(z)+\sigma V_{11}(z)&\approx&
-\frac{\Lambda_0^A}{z}+(\Lambda_0^B+\sigma \Lambda^B_{1})
\left(z+\sigma \frac{k-1}{6} z^{2}\right).
\nonumber
\end{eqnarray}

\section{Matching between Inner and Outer Solutions}\label{sematch}

\subsection{Matching at the singular point $t=t_1$}
In this section, solutions of the outer equations are constructed
and they are matched with the inner solutions.
We first consider the contribution from the singularity $t=t_1$.
Corresponding to the expansion of the analytically continued inner solutions:
$V=V_0+e^{-2\pi i z}V_1+\cdots$, $U= U_0+e^{-2\pi i z}U_1+\cdots$, 
the original solutions $v$ and $u$ acquire new terms in a sector ${\rm Re} \ t\ge {\rm Re}\ t_1=0$ of the 
$t_1$-neighborhood:
\begin{eqnarray}
v(t)=v_{0}(t,\sigma)+v_{1}(t,\sigma)e^{-\frac{2\pi i}{\sigma}t}+ \cdots 
\nonumber
\\
u(t)=u_{0}(t,\sigma)+u_{1}(t,\sigma)e^{-\frac{2\pi i}{\sigma}t}+ \cdots
\label{original new exp}
\end{eqnarray}
Because of $e^{-2\pi i z}= e^{-{\pi^2 \over \sigma \sqrt{k}}} e^{-{2\pi i \over \sigma}t}$,
$v_1$ and $u_1$ are exponentially small with respect to $\sigma$.
By substituting  (\ref{original new exp}) into 
(\ref{eq:Harper}) and comparing term by term,
we obtain the equations for $v_1$ and $u_1$:
\begin{eqnarray}
v_{1}(t+\sigma)-v_{1}(t) &=&-\sigma u_{1}(t) \cos u_{0}(t)
\nonumber
\\
u_{1}(t+\sigma)-u_{1}(t) &=&k\sigma v_{1}(t+\sigma) \cos v_{0}(t+\sigma)
\label{Harper1}
\end{eqnarray}

Its solution is uniquely determined by the matching condition:
\begin{eqnarray}
\left. e^{-\frac{2\pi it}{\sigma}}
\pmat{
v_{1}(t,\sigma)
\cr
u_{1}(t,\sigma)
}
\right|_{t=t_1+\sigma z}
&=& e^{-2\pi iz}
\pmat{ V_{10}(z) +\sigma V_{11}(z)+ \cdots \cr 
U_{10}(z) +\sigma U_{11}(z) +\cdots } 
\end{eqnarray}
or equivalently
\begin{eqnarray}
\pmat{
v_{1}(t_1+\sigma z,\sigma)
\cr
u_{1}(t_1+\sigma z,\sigma)
}
&=& e^{\frac{2\pi it_1}{\sigma}}
\pmat{ V_{10}(z) +\sigma V_{11}(z)+ \cdots \cr 
U_{10}(z) +\sigma U_{11}(z) +\cdots } 
\label{matchrel}
\end{eqnarray}
This condition suggests the following expansions:
\begin{eqnarray}
v_1(t,\sigma)=\sigma^j e^{\frac{2\pi it_1}{\sigma}}\sig{n}v_{1n}(t)\sigma^n
\ , \qquad
u_1(t,\sigma)=\sigma^j e^{\frac{2\pi it_1}{\sigma}}\sig{n}u_{1n}(t)\sigma^n
\end{eqnarray}
where the exponent $j$ will be determined later.
Then, equations for $v_{10},u_{10},v_{11},u_{11}$ are given by
\begin{eqnarray}
&&v'_{10}(t)=-u_{10}(t)\cos u_{00}(t)
\nonumber
\\
&&u'_{10}(t)=kv_{10}(t)\cos v_{00}(t)
\label{ODE:10}
\\
&&v'_{11}(t)+\frac{1}{2}v''_{10}(t)=-u_{11}(t)\cos u_{00}(t)+u_{10}(t)u_{01}(t)\sin u_{00}(t)
\nonumber
\\
&&u'_{11}(t)-\frac{1}{2}u''_{10}(t)=k\left(v_{11}(t)\cos v_{00}(t)-v_{10}(t)v_{01}(t)\sin v_{00}(t)\right)
\label{ODE:11}
\end{eqnarray}

First, we solve (\ref{ODE:10}). Since it is linear, one has
\begin{eqnarray}
\pmat{
v_{10}(t)\cr u_{10}(t)
}
&=&
a
\pmat{x_1(t)\cr y_1(t)
}
+
b
\pmat{x_2(t) \cr y_2(t)
}
\end{eqnarray}
where $x_1(t), y_1(t), x_2(t), y_2(t)$ are defined by 
(\ref{eq:x1}) and (\ref{eq:x2}).

On the other hand, 
up to the order of $\sigma^0$, the matching condition leads to
\begin{eqnarray}
\sigma^j
\left.
\pmat{
v_{10}(\sigma z +t_1)
\cr
u_{10}(\sigma z +t_1)
}
\right|_{\sigma^0}
&=&
\sigma^j 
\left.
\left\{
a
\pmat{x_1(\sigma z +t_1)
\cr	
y_1(\sigma z +t_1)}
+
b
\pmat{x_2(\sigma z +t_1)
\cr	
y_2(\sigma z +t_1)
}
\right\}
\right|_{\sigma^0}
\nonumber \\
&\approx&
\left.
\pmat{ V_{10}(z) +\sigma V_{11}(z)+ \cdots \cr 
U_{10}(z) +\sigma U_{11}(z) +\cdots } 
\right|_{\rm dom.}
=
\Lambda_0^B
\pmat{
z
\cr	
z
}
\label{apprMatch}
\end{eqnarray} 
where the subscript $\sigma^0$ indicates to take the terms of order $\sigma^0$ 
and `dom.' stands for the largest part for $z\to \infty$.
Because the left hand side of (\ref{apprMatch}) starts from $z$, one should have $j=-1$.
And the coefficients $a$ and $b$ are determined by the requirement that
(\ref{apprMatch}) admits well defined $\sigma\to 0$ limit:
$$
a=0
\ , \qquad
b=\frac{2\Lambda_0^B}{i} \ .
$$
Therefore, up to the $0$th order in $\sigma$, we have
\begin{eqnarray}
\pmat{
v_{1}(t)
\cr
u_{1}(t)
}
\approx
\frac{2\Lambda_0^B}{i\sigma}
e^{-\frac{2\pi i(t-t_{1})}{\sigma}}
\pmat{
x_{2}(t)
\\ 
y_{2}(t)
}
\label{ac}
\end{eqnarray}
Similarly, up to the first order in $\sigma$, one obtains
\begin{eqnarray}
\pmat{
v_{1}(t)
\\ 
u_{1}(t)
}
&\approx&
\frac{2\Lambda^{(1)}}{i\sigma}
e^{-\frac{2\pi i(t-t_{1})}{\sigma}}
\pmat{
x_{2}(t)
\\ 
y_{2}(t)+\sigma y'_{2}(t)/2
}
\label{af}
\end{eqnarray}
where $\Lambda^{(1)}\equiv\Lambda_0^B+\sigma\Lambda_1^B$.
Matching of higher order terms with respect to $\sigma$ is discussed in 
Appendix~\ref{GeneralMatch}.

\subsection{Contribution from the singular point at $t=t_1^*$}

So far, the behavior of the solution near $t=t_1$ has been considered.
In order to derive the solution valid in the domain Re~$t>0$, the contribution from the
complex conjugate singularity $t=t_1^*$ should be taken into account.
Because $v_0(t), u_0(t)$ is real analytic,
it is simply the complex conjugate of the contribution from $t_1$ 
(See appendix~\ref{app i.c.}).
Hence, the unstable manifold for $t<2T$ is described by
\begin{eqnarray}
\pmat{v_{u}(t) \\ u_{u}(t)}=
\pmat{
v_{00}(t)+\sigma^2 v_{02}(t) \\
u_{00}(t)+\sigma \frac{y_{1}(t)}{2}+\sigma^{2}u_{02}(t)+\sigma^{3} 
\left(\frac{1}{2}u'_{02}(t)-\frac{1}{24}y''_{1}(t)\right)
}
\nonumber \\
~+S(t)
{\rm Re}\left[\frac{4\Lambda^{(1)}}{i\sigma}
e^{-\frac{2\pi i(t-t_{1})}{\sigma}}
\pmat{
x_{2}(t)
\\ 
y_{2}(t)+\sigma y'_{2}(t)/2
}
\right]
\label{onlyone}
\end{eqnarray}
where $S(t)$ denotes a step function.

\section{Contributions from Other Singularities}\label{sec:other}
So far, contributions from $t_1$ and $t_1^*$ have been considered.
Here, we study contributions from other singular points and higher order terms.
As easily seen, in the domain Re~$t>0$, the expression (\ref{onlyone}) of the unstable manifold 
has a singularity at 
$$
t \equiv t_2= 2T+\frac{1}{\sqrt{k}}\left(n+\frac{1}{2}\right)\pi i
\ .
$$
Therefore, the solution would change its form when $t$ exceeds $2T$. As in the previous analysis, the additional terms can be
found from the asymptotic behavior of the solution near $t_2$ 
and its complex conjugate $t_2^*$. 

The solution (\ref{onlyone}) valid in $t<2T$ is a sum of the perturbative 
solution and the terms proportional to $S(t)$, and 
each term produces a new additional term in the
sector Re~$t >2T$. 
We analyze these two terms separately.
\begin{itemize}
\item[(1)]
\underline{The additional terms arising from the perturbative solution.}
\\
The additional terms arising from the perturbative solution 
can be obtained exactly in the same way as those from $t=t_1$ and we have
\begin{eqnarray}
\pmat{V^-_{01}(z) \cr U^-_{01}(z)}=
i\left\{
\frac{kt_2 -1}{24}\pmat{V_{A}(z) \cr -U_{A}(z)}+
\frac{k-1}{2}\pmat{-(V_{B}(z)-z) \cr U_{B}(z)-(z-\frac{1}{2})}
\right\}
\end{eqnarray}

\begin{eqnarray}
\pmat{ v_{1}^{(2)}(t)
\cr u_{1}^{(2)}(t)}
=
{\rm Re}\biggl[
\frac{4\Lambda^{(2)}}{i\sigma}
e^{-\frac{2\pi i(t-t_2)}{\sigma}}
&\Big\{&
\frac{T(k-1)^2-(1+k)}{4k(k-1)}\pmat{
x_{1}(t)
\cr 
y_{1}(t)+\sigma y'_{1}(t)/2
}\nonumber\\ 
&&+\pmat{
x_{2}(t)
\cr 
y_{2}(t)+\sigma y'_{2}(t)/2
}
\Big\}
\biggr], \  \
\label{second}
\end{eqnarray}
where $\Lambda^{(2)}$ is given by 
$\Lambda^{(2)}\equiv -i4\pi^{3}A_{1}+
\sigma\pi^3\left\{
-(k-1)B_2+\frac{kt_2-1}{12}B_4
\right\}$. 

\item[(2)] \underline{The additional terms arising from the 
term proportional to $S(t)$}. \\
Let us prove that the terms added in the domain $0< {\rm Re} t<2T$
generate new terms which are of order of $\epsilon^2$ and, thus, 
are negligible.
With the same analysis as for the most divergent terms from the 
perturbative solution, their most divergent terms near $t=t_2$ are found to be
\begin{eqnarray}
\frac{i\Gamma}{\sigma}
\xi (t)\pmat{V_A(z') \cr -U_A(z')}
-\frac{i\Gamma \sigma}{2}\cos\frac{2\pi t}{\sigma}\pmat{V_B(z') \cr -U_B(z')},
\label{additinDiv}
\end{eqnarray}
where $\xi(t), \Gamma$ are defined by
\begin{eqnarray}
\xi (t)&\equiv&-\frac{t_2(1-k)^2-2(1+k)}{8k(k-1)}\cos\frac{2\pi t}{\sigma}
+\frac{t_1(k-1)}{8ik}\sin\frac{2\pi t}{\sigma}
\nonumber \\
\Gamma&\equiv&\frac{4\Lambda^B_0}{i\sigma}e^{-\frac{\pi^2}{\sqrt{k}\sigma}}
\end{eqnarray}
This can be represented with the aid of the Borel transformations:
\begin{eqnarray*}
\frac{i\Gamma}{\sigma}
\xi (t)\int^{-\infty}_{0} dp \hspace*{1mm} e^{-pz'}
\pmat{\widetilde{V}_A(p) \cr -\widetilde{U}_A(p)}
-\frac{i\Gamma \sigma}{2}\cos\frac{2\pi t}{\sigma}
\left[\int^{-\infty}_{0} dp \hspace*{1mm} e^{-pz'}
\pmat{\widetilde{V}_B(p) \cr -\widetilde{U}_B(p)}
+\pmat{z' \cr z'+1/2}
\right]
\end{eqnarray*}
where $(\widetilde{V}_A(p),\widetilde{U}_A(p))$ and 
$(\widetilde{V}_B(p),\widetilde{U}_B(p))$ are, respectively, 
the Borel transforms of
$(V_A(z'),U_A(z'))$ and $(V_B(z')-z',U_B(z')-z'-1/2)$.
Thus the following term is added in Re$t>2T$
\begin{eqnarray*}
&&-\frac{i\Gamma}{\sigma}
\xi (t)\int_{\gamma} dp \hspace*{1mm} e^{-pz'}
\pmat{\widetilde{V}_A(p) \cr -\widetilde{U}_A(p)}
+\frac{i\Gamma \sigma}{2}\cos\frac{2\pi t}{\sigma}
\int_{\gamma} dp \hspace*{1mm} e^{-pz'}
\pmat{\widetilde{V}_B(p) \cr -\widetilde{U}_B(p)}
\\
=
&&\Gamma e^{-2\pi iz'}\Bigg[
\frac{2\pi^3 B_4 z'}{\sigma}\xi (t)
\pmat{1 \cr -1}
\\&&+\frac{i \sigma}{2}\cos\frac{2\pi t}{\sigma}
\left\{
8\pi^4 B_1 z'^2\pmat{1 \cr 1}
+2\pi^3 i B_2 z'\pmat{1 \cr -1}
-2\pi^3 B_3 z'\pmat{0 \cr -1}
\right\} \Bigg]
\\
\simeq
&&\Gamma e^{-2\pi iz'}\Bigg[
\frac{2\pi^3 B_4 (t-t_2)}{\sigma^2}\xi (t)
\pmat{1 \cr -1}
+\frac{i }{2\sigma}(t-t_2)^2\cos\frac{2\pi t}{\sigma}
8\pi^4 B_1 \pmat{1 \cr 1}
\Bigg],
\end{eqnarray*}
which leads to terms of order of 
$\frac{1}{\sigma^3}e^{-\frac{2\pi^2}{\sqrt{k}\sigma}}$
in the outer solution.
Therefore this term is negligible when 
$\displaystyle{\frac{1}{\sigma^3}e^{-\frac{2\pi^2}{\sqrt{k}\sigma}}\ll 
\frac{1}{\sigma}e^{-\frac{\pi^2}{\sqrt{k}\sigma}}}$.
\end{itemize}
As a result, the overall contribution 
of the singular points is the simple sum of the contributions 
from $t_1,\ t_2,\ t_1^*,\ t_2^*$.
Because (\ref{second}) has no singularity in the domain Re~$t>2T$, 
one finally has the following expression of the unstable manifold valid in the whole time domain:
\begin{eqnarray}
v_{u}(t)&&=
v_{00}(t)+\sigma^2 v_{02}(t) 
\nonumber \\
&&~+S(t)
{\rm Re}\left[
\frac{4\Lambda^{(1)}}{i\sigma}
e^{\frac{2\pi it_{1}}{\sigma}}
x_2(t)
e^{-\frac{2\pi it}{\sigma}}
\right]
\nonumber
\\
&&~+S_+(t)
{\rm Re}\left[
\frac{4\Lambda^{(2)}}{i\sigma}
e^{\frac{2\pi it_{2}}{\sigma}}
\left\{
\left(\frac{T(k-1)^2-(1+k)}{4k(k-1)}\right)x_{1}(t)+x_2(t) 
\right\}
e^{-\frac{2\pi it}{\sigma}}
\right]
\nonumber
\\ \nonumber \\
u_{u}(t)
&&=
u_{00}(t)+\sigma \frac{y_{1}(t)}{2}+\sigma^{2}u_{02}(t)+\sigma^{3} 
\left(\frac{1}{2}u'_{02}(t)-\frac{1}{24}y''_{1}(t)\right)
\nonumber
\\
&&~+S(t)
{\rm Re}\biggl[
\frac{4\Lambda^{(1)}}{i\sigma}
e^{\frac{2\pi it_{1}}{\sigma}}
\left(y_{2}(t)+\sigma\frac{y'_{2}(t)}{2}\right)
e^{-\frac{2\pi it}{\sigma}}
\biggr]
\nonumber
\\
&&~+S_+(t)
{\rm Re}\biggl[
\frac{4\Lambda^{(2)}}{i\sigma}
e^{\frac{2\pi it_{2}}{\sigma}}
\biggl\{ \left(\frac{T(k-1)^2-(1+k)}{4k(k-1)}\right)
\left(y_{1}(t)+\sigma\frac{y'_{1}(t)}{2}\right)
\nonumber \\
&& \mskip 320 mu +
\left(y_{2}(t)+\sigma\frac{y'_{2}(t)}{2}\right) \biggr\}
e^{-\frac{2\pi it}{\sigma}}
\biggr]
\label{final unstable man}
\end{eqnarray}
where $S_+(t)=S(t- 2T)$ and $S(t)$ stands for 
the step function, and $\Lambda^{(1)}$,\ $\Lambda^{(2)}$ are given previously.
Before closing this section, we give an approximate expression of the stable manifold
$v_s(t), u_s(t)$, which is the time-reversal image of the unstable manifold:
\begin{eqnarray}
v_{s}(t)&&=
v_{00}(t)+\sigma^2 v_{02}(t) 
\nonumber \\
&&~-S(-t)
{\rm Re}\left[
\frac{4\Lambda^{(1)}}{i\sigma}
e^{\frac{2\pi it_{1}}{\sigma}}
x_2(t)
e^{-\frac{2\pi it}{\sigma}}
\right]
\nonumber
\\
&&~-S_+(-t)
{\rm Re}\left[
\frac{4\Lambda^{(2)}}{i\sigma}
e^{\frac{2\pi it_{2}}{\sigma}}
\left\{
\left(\frac{T(k-1)^2-(1+k)}{4k(k-1)}\right)x_{1}(t)+x_2(t) 
\right\}
e^{-\frac{2\pi it}{\sigma}}
\right]
\nonumber
\\ \nonumber \\
u_{s}(t)
&&=
u_{00}(t)+\sigma \frac{y_{1}(t)}{2}+\sigma^{2}u_{02}(t)+\sigma^{3} 
\left(\frac{1}{2}u'_{02}(t)-\frac{1}{24}y''_{1}(t)\right)
\nonumber
\\
&&~-S(-t)
{\rm Re}\biggl[
\frac{4\Lambda^{(1)}}{i\sigma}
e^{\frac{2\pi it_{1}}{\sigma}}
\left(y_{2}(t)+\sigma\frac{y'_{2}(t)}{2}\right)
e^{-\frac{2\pi it}{\sigma}}
\biggr]
\nonumber
\\
&&~-S_+(-t)
{\rm Re}\biggl[
\frac{4\Lambda^{(2)}}{i\sigma}
e^{\frac{2\pi it_{2}}{\sigma}}
\biggl\{ \left(\frac{T(k-1)^2-(1+k)}{4k(k-1)}\right)
\left(y_{1}(t)+\sigma\frac{y'_{1}(t)}{2}\right)
\nonumber \\
&& \mskip 300 mu +
\left(y_{2}(t)+\sigma\frac{y'_{2}(t)}{2}\right) \biggr\}
e^{-\frac{2\pi it}{\sigma}}
\biggr]
\label{final ap}
\end{eqnarray}

\begin{figure}[htbp]
\begin{center}
\includegraphics[width=8cm,keepaspectratio,clip]{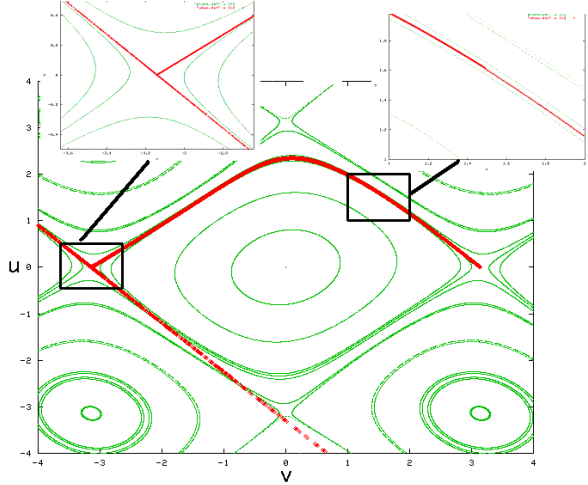}
\caption{The analytically constructed unstable manifold (solid line)
and numerically portrated phase space.\ ($\sigma=0.35,k=0.85$)}
\label{overal}
\end{center}
\end{figure}

\begin{figure}[htbp]
\begin{tabular}{cc}
\rotatebox[origin=c]{-90}{
\includegraphics[scale=0.26]{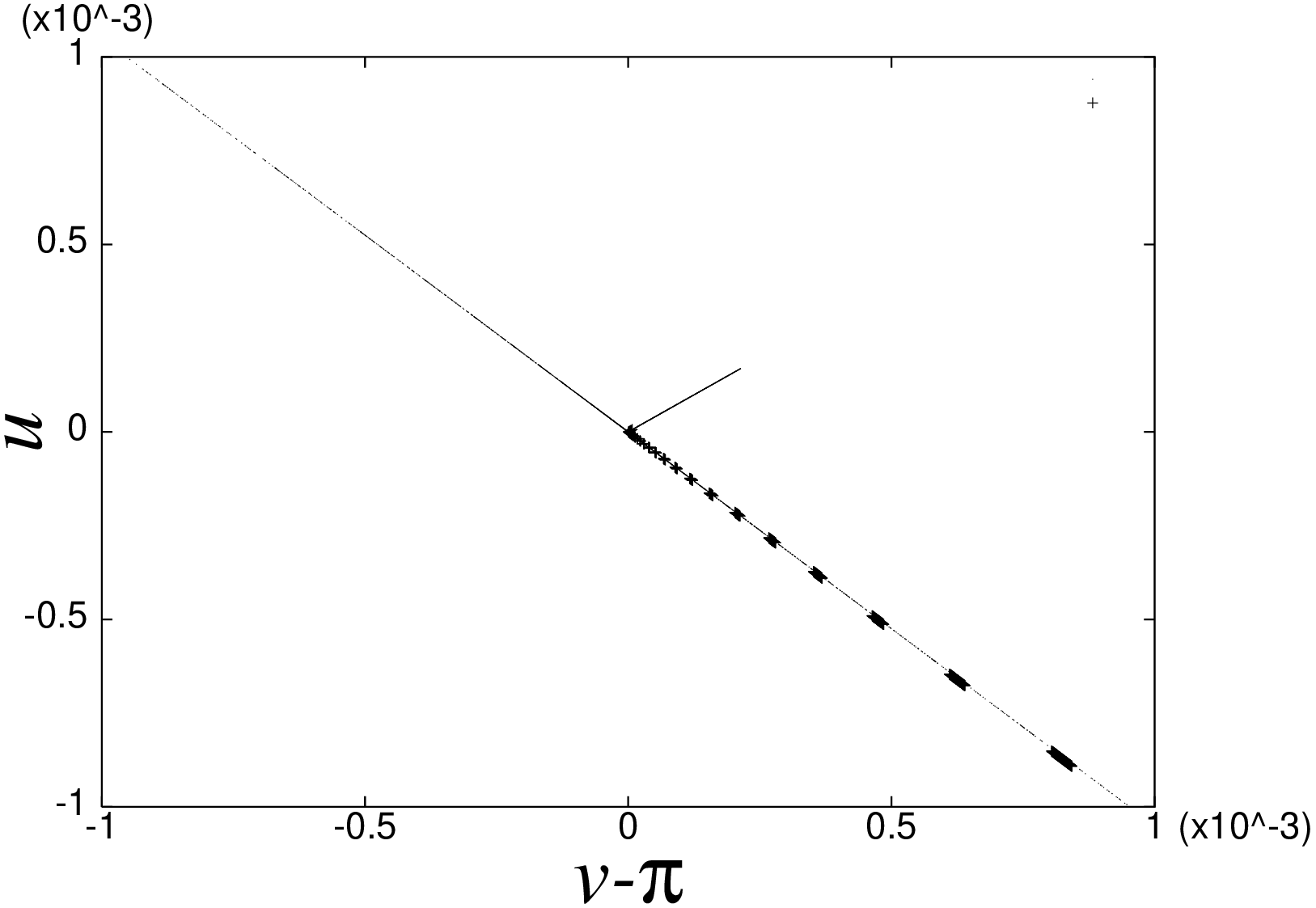}}&
\rotatebox[origin=c]{-90}{
\includegraphics[scale=0.26]{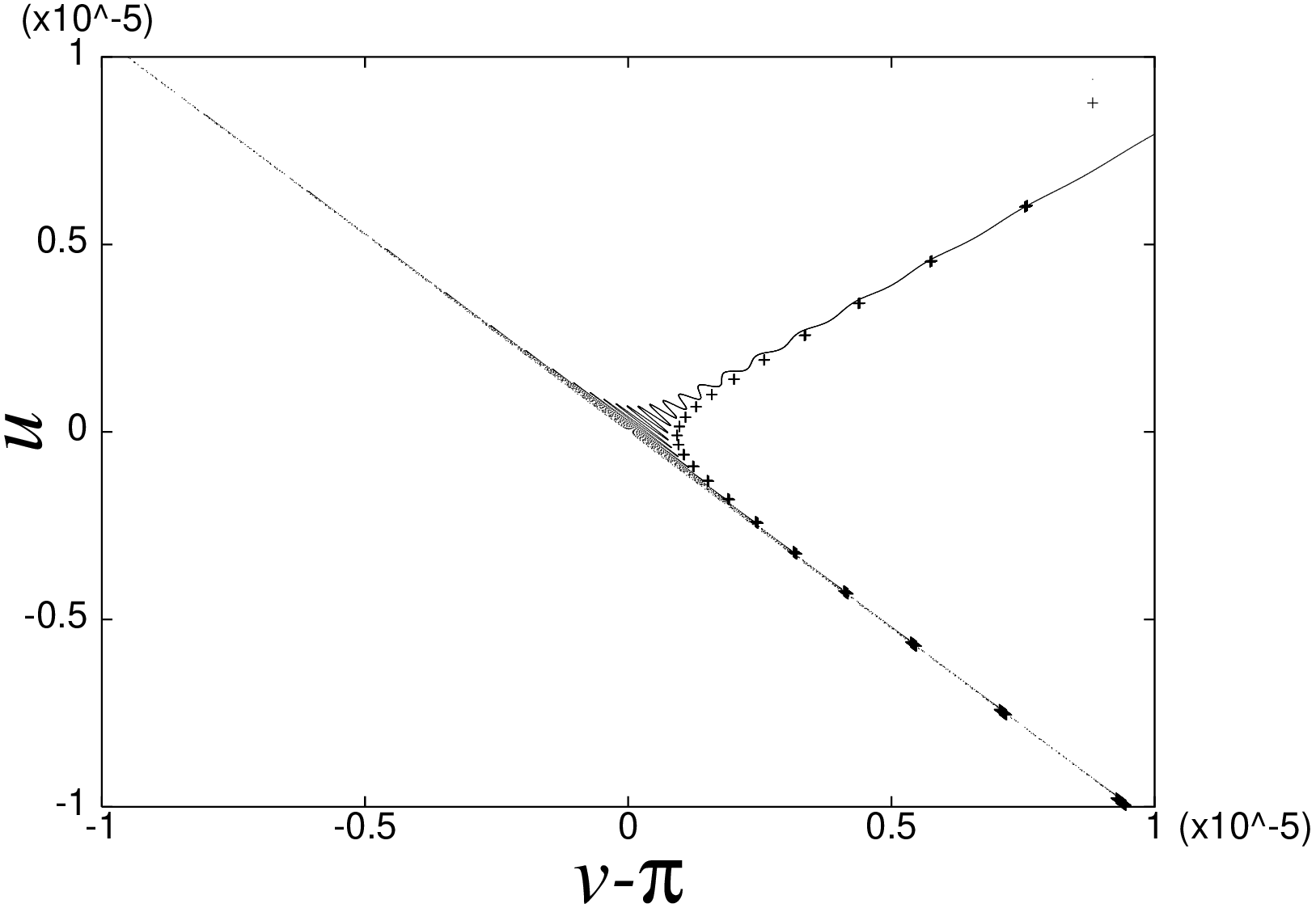}}
\end{tabular}
\caption{The analytically constructed unstable manifold (solid line)
near $(-\pi,0)$ and 
time evolution of the ensemble (plus-mark).\ ($\sigma=0.3,k=0.85$)}
\label{full}
\end{figure}

\begin{figure}[htbp]
\begin{tabular}{cc}
\rotatebox[origin=c]{-90}{
\includegraphics[scale=0.26]{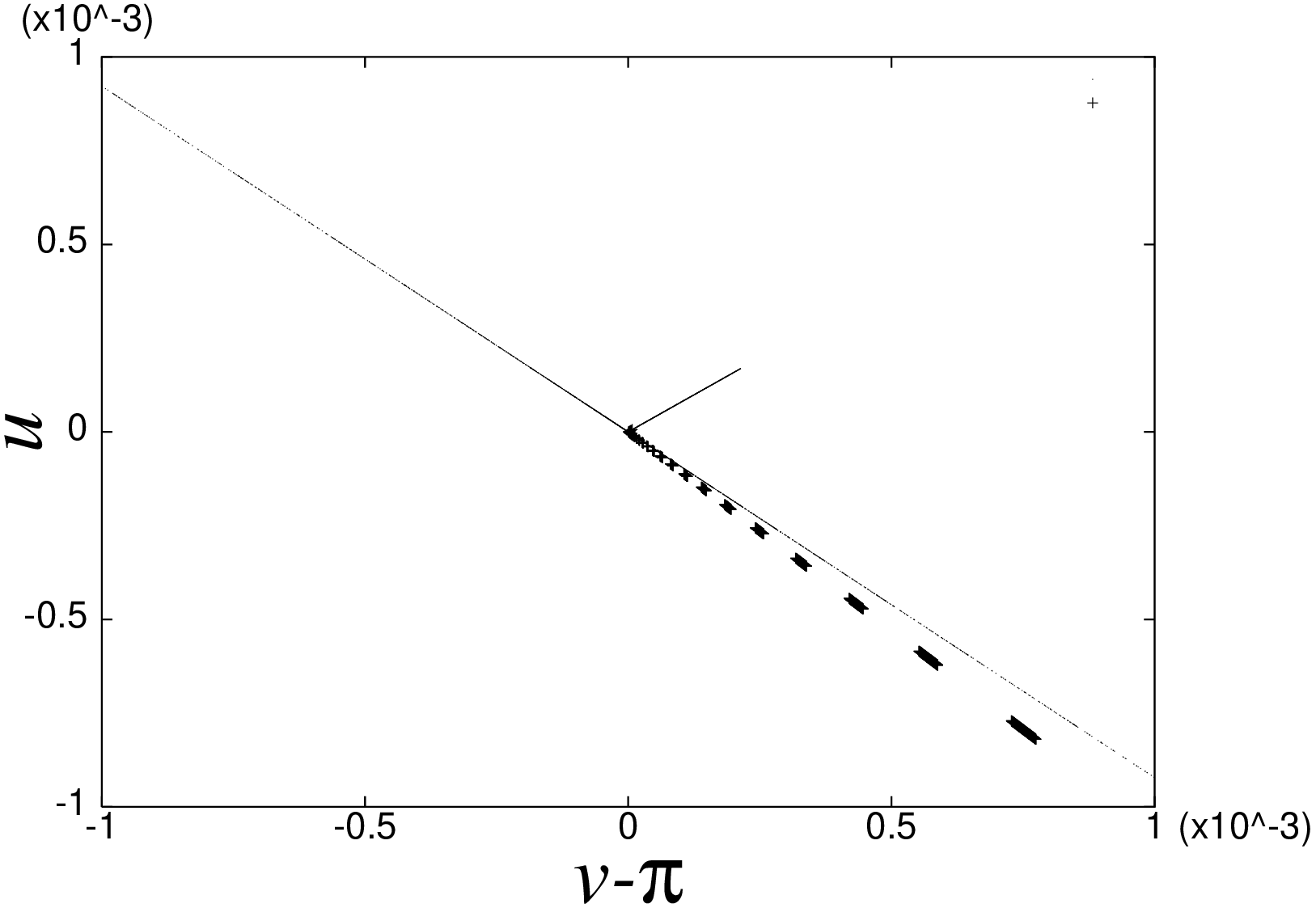}}&
\rotatebox[origin=c]{-90}{
\includegraphics[scale=0.26]{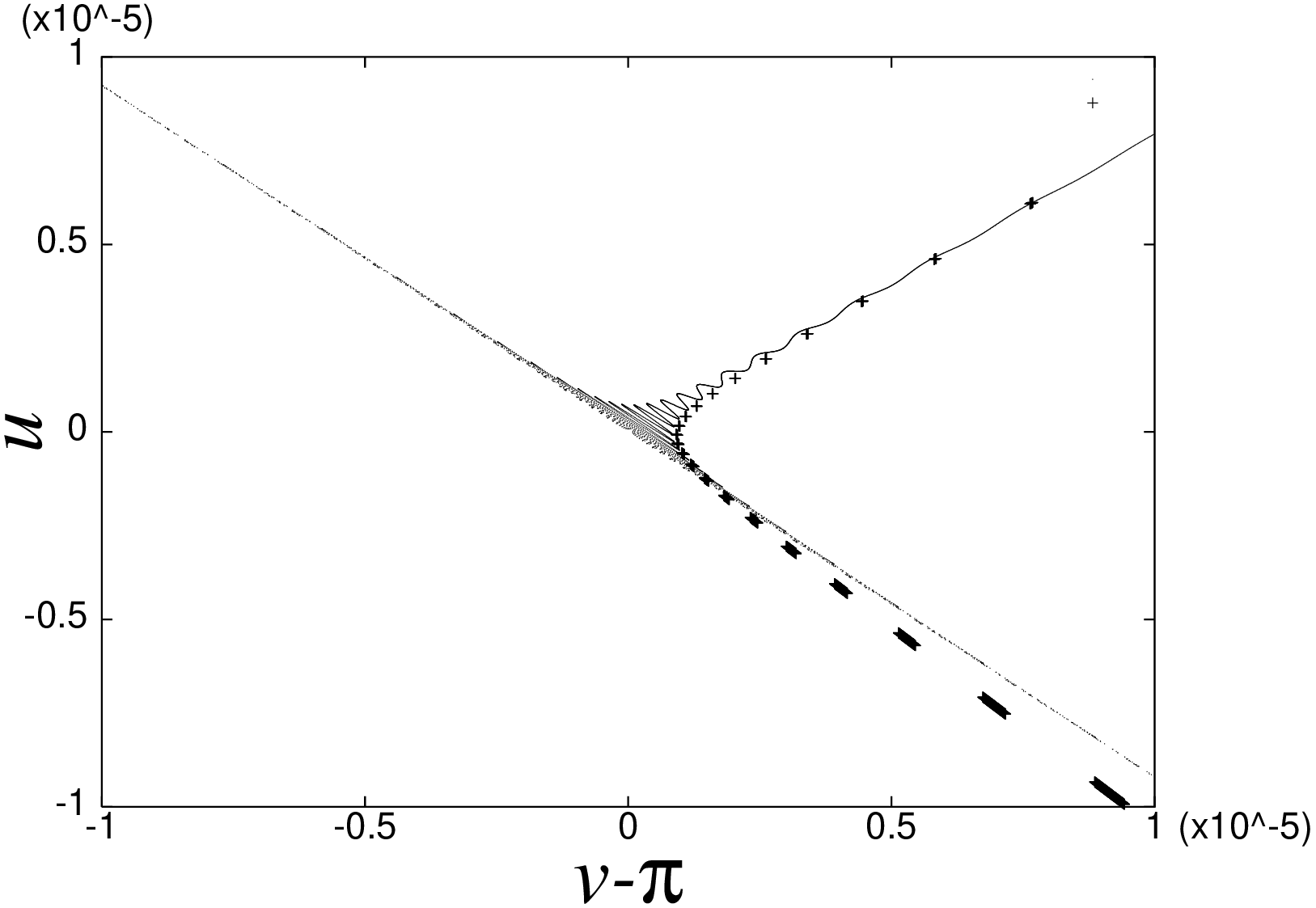}}
\end{tabular}
\caption{The analytically constructed unstable manifold (solid line)
near $(-\pi,0)$ and 
time evolution of the ensemble (plus-mark).\ 
Only the leading term is taken into consideration.
($\sigma=0.3,k=0.85$)}
\label{dominant}
\end{figure}

\begin{figure}[htbp]
\begin{tabular}{cc}
\rotatebox[origin=c]{-90}{
\includegraphics[scale=0.26]{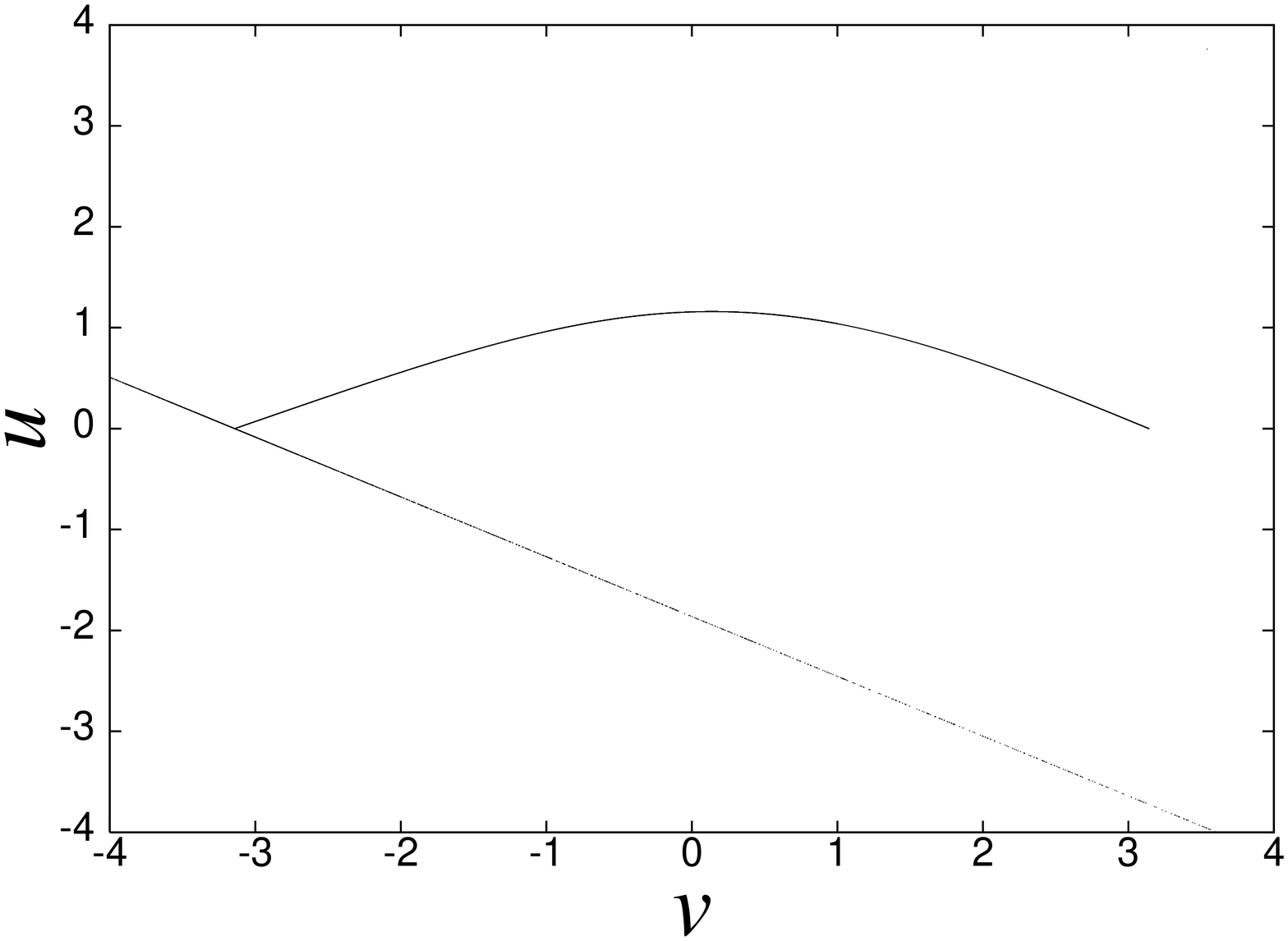}}&
\rotatebox[origin=c]{-90}{
\includegraphics[scale=0.26]{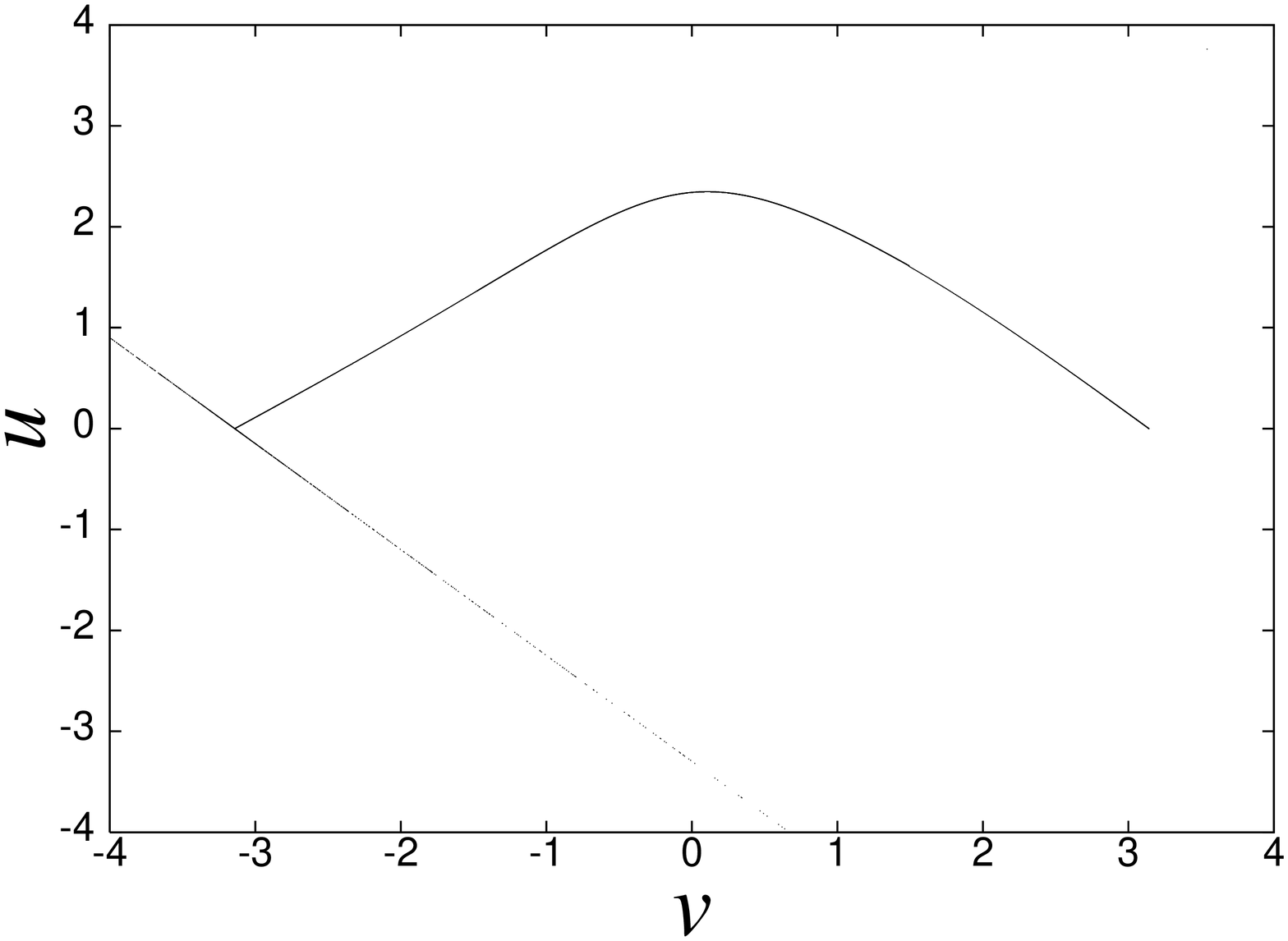}}
\end{tabular}
\caption{The analytically constructed unstable manifold.
The left figure shows the case of $\sigma=0.3,\ k=0.3$ and the right figure shows the case of $\sigma=0.3,\ k=0.85$.}
\label{process1}
\end{figure}

\begin{figure}[htbp]
\begin{center}
\rotatebox[origin=c]{-90}{
\includegraphics[width=8cm,keepaspectratio,clip]{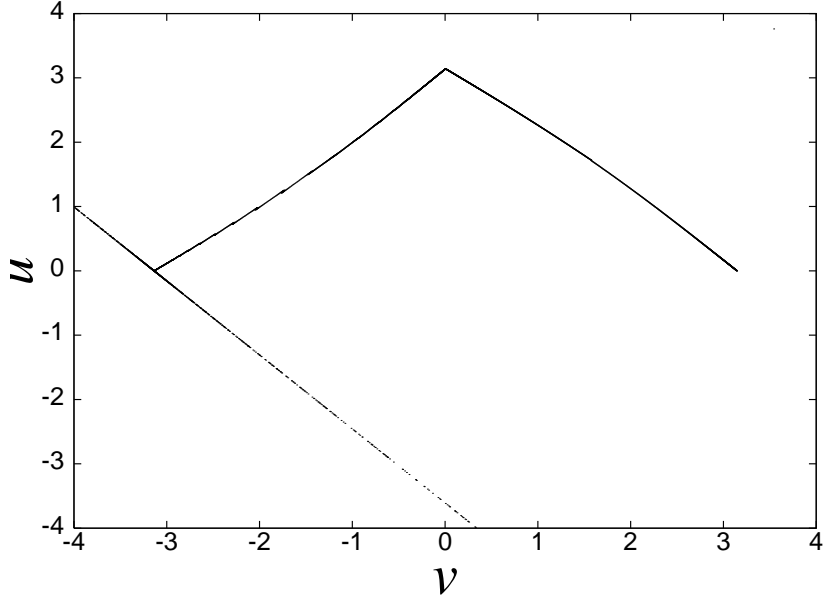}}
\caption{The analytically constructed unstable manifold for 
$\sigma=0.3,\ k=1.0-5.0\times 10^{-11}$.}
\label{process2}
\end{center}
\end{figure}

\begin{figure}[htbp]
\begin{tabular}{cc}
\rotatebox[origin=c]{-90}{
\includegraphics[scale=0.26]{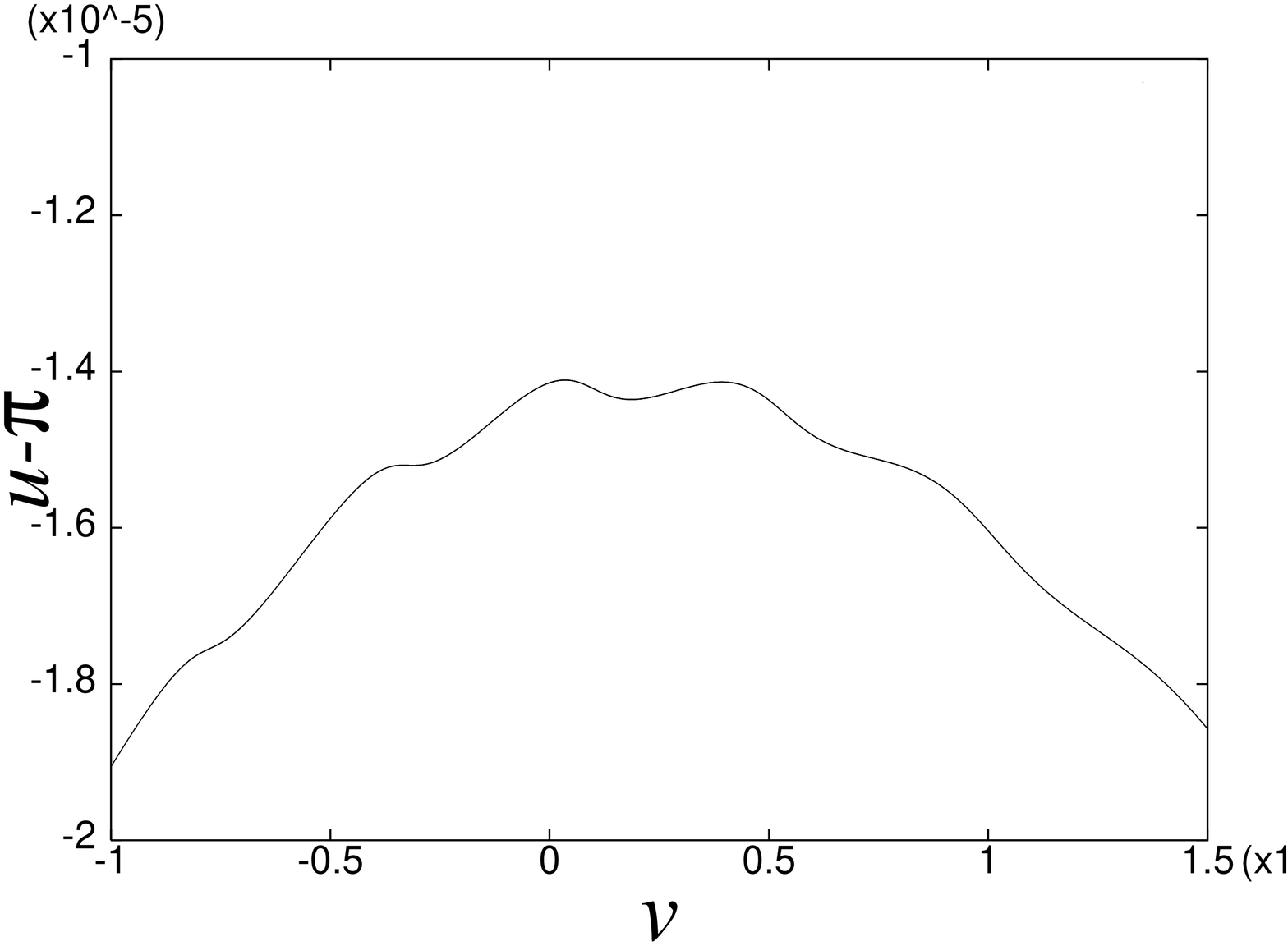}}&
\rotatebox[origin=c]{-90}{
\includegraphics[scale=0.26]{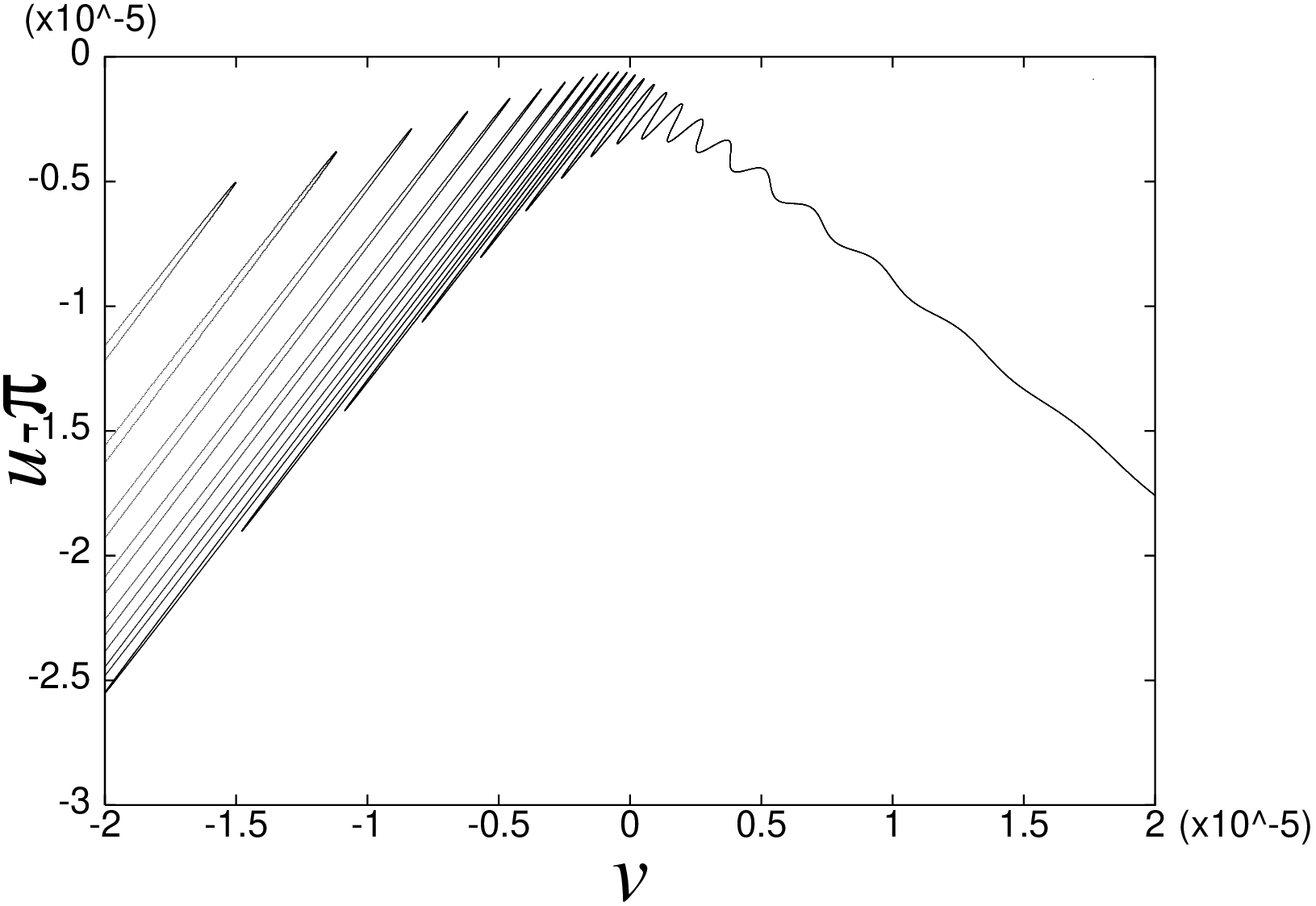}}
\end{tabular}
\caption{The analytically constructed unstable manifold near $(0,\pi)$.
The left figure shows the case of $\sigma=0.3,\ k=1.0-5.0 \times 10^{-11}$ and 
the right figure shows the case of $\sigma=0.3,\ k=1.0-1.4\times 10^{-12}$.}
\label{reconnection}
\end{figure}

\section{Comparison with Simulations and 
Reconnection of Unstable Manifold}
\subsection{Comparison with numerical calculations}

Here we compare the analytical solution (\ref{final unstable man}) obtained in 
the previous section with the numerical calculation. 
(\ref{final unstable man}) indicates that, when $t$ exceeds $t=0$, 
exponentially growing oscillatory term appears and, when $t$ exceeds $t=2T$,
another exponentially growing term is added. These terms 
are exponentially small with respect to $\sigma$ 
but they grow exponentially and
become dominant for sufficiently large $t$.
Thus, added terms describe heteroclinic tangles and
a large-amplitude oscillation appears near $(-\pi,0)$ can be explained with
them.
This is indeed the case. Fig.~\ref{overal} shows
the analytically obtained approximate solution of 
the unstable manifold from $(\pi,0)$ (solid curve)
and numerically portrayed phase space, 
and \ Fig.~\ref{full} shows 
the analytically obtained approximate solution of the unstable manifold from $(\pi,0)$ (solid curve) 
and the numerically calculated time evolution of 
an ensemble of 500 points on solid curve (plus-mark) near $(-\pi,0)$.
As shown in Fig.~\ref{full},
the analytically obtained approximate solution 
(\ref{final unstable man}) (solid curve) well reproduces the 
numerical results (plus-mark), even the large-amplitude oscillation and 
stretching of the ensemble near $(-\pi,0)$. 
When only the leading order terms of the inner equation are retained, the agreement 
between the analytical and numerical 
results is not so good (cf. Fig.~\ref{dominant}).
We also observe that terms of order $\sigma$ extends time range in which 
the solution well represents the unstable manifold and, thus, 
the terms of order $\sigma$ of the inner solution are important.

Fig.~\ref{process1},\ Fig.~\ref{process2} 
show the unstable manifolds when $\sigma=0.30$ and 
$k=0.3,\ 0.85,\ 1.0-5.0\times 10^{-11}$, respectively.
The overall structure of the unstable manifold is similar to that of 
the separatrix except near the fixed point $(-\pi,0)$.
Near $(-\pi,0)$, the unstable manifold acquires an oscillatory portion
and the slope of this portion becomes steeper as $k$ increases 1.
The behavior of the slope can be understood from the asymptotic ratio 
between the additional terms:
$$
\displaystyle \lim_{t\to +\infty}\frac{u_{1}(t)}{v_{1}(t)+\pi}=-\sqrt{k}-\frac{k}{2}\sigma
\ .
$$
\noindent In Fig.~\ref{process1},\ Fig.~\ref{process2} one can see a small stochastic domain and this ratio represent an unstable direction in the domain near $(-\pi,0)$.

\subsection{Reconnection transition}

We remind that the separatrix of the continuous-time limit equation is the unperturbed unstable/stable 
manifold.  As mentioned in introduction (Fig.~\ref{fig:differential}), 
the separatrix changes its topology depending on the parameter $k$
(reconnection). When $0< k <1$, there appears a separatrix connecting
$(\pi,0)$ and $(-\pi,0)$. 
As $k\to 1$, the time when the orbit crosses the $u$-axis diverges and
it approaches the separatrix for $k=1$ connecting $(\pi,0)$ with $(0,\pi)$. 
The corresponding topological change occurs for unstable/stable manifolds,
which will be discussed in subthis section.
As shown in the previous section, since the approximate stable and 
unstable manifolds are related with each other by a simple symmetry,
it is enough to discuss the topological change of the unstable
manifold.
It is interesting to see that, when $k$ is very close to unity (i.e., 
$k=1.0-5.0\times 10^{-11}$), 
there appears 
an additional oscillatory portion in the unstable manifold near $(0,\pi)$
(cf. Fig.~\ref{process2}, Fig.~\ref{reconnection}).
As mentioned before, at $k=1$, two segments from ($\pi,0$) to ($0,\pi$) and 
from ($0,\pi$) to ($-\pi,0$)
are separatrices of the continuous-limit equation.
Thus, the perturbed unstable manifold at $k=1$ starting from $(\pi,0)$ should have an
oscillatory portion near $(0,\pi)$. The oscillation near $(0,\pi)$ of the unstable manifold 
shown in Fig.~\ref{reconnection} can be considered as the precursor of this oscillation
in the unstable manifold at $k=1$.
Origins of these behaviors are summarized as follows.

When $k$ is not very close to 1, the unstable manifold reaches near $(-\pi,0)$ 
before additional terms become large.
On the other hand, when $k$ is very close to 1, the orbit along the unstable manifold takes very long time to reach $u$-axis because $v_{00}(T(k))=0$ and 
$\displaystyle{\lim_{k\to 1-0}T(k)=\infty}$.
Moreover, since the solutions at $t=T(k)$ behave:
\begin{eqnarray*}
\lim_{k\to 1-0}\pmat{v_{00}(T) \cr u_{00}(T)}&=&
\pmat{0 \cr \pi}\\
\lim_{k\to 1-0}\pmat{x_2(T) \cr y_2(T)}&=&
\pmat{\infty \cr -\infty},
\end{eqnarray*}
when $k\simeq 1$ the additinal term 
is not so small compared to the perturbative solution near $(0,\pi)$. 
Therefore the unstable manifold shows a small oscillation near $(0,\pi)$.
We note that the oscillation near $(0,\pi)$ is driven by the first 
additinal term, on the contrary, the oscillation near $(-\pi,0)$ is driven by 
the two additinal terms.
We further remark that two oscillatory portions in the unstable manifold
for $k\simeq 1$ come from the singularities at $t=t_1,t_1^*$ and the singularities 
$t=t_2,t_2^*$ goes to infinity as 
$\displaystyle{\lim_{k\to 1-0} {\rm Re}~t_2=\infty}$.
In other words, the two sequences of singularities are necessary for the description of
the reconnection of unstable/stable manifolds.

In Fig.~\ref{reconnection}, the unstable manifold for 
$k=1.0-5.0\times 10^{-11}$ and $k=1.0-1.4\times 10^{-12}$ 
near $(0,\pi)$ are shown.
Because of the exponential divergence ($\sim e^{\sqrt{k}t}$) of the additional terms for $t\to\infty$, the approximate unstable 
manifold (\ref{final unstable man})
will not work if $|\epsilon e^{\sqrt{k}t}|\gg 1$, i.e. 
$t\gg \frac{\pi^2}{k\sigma}$.
When $k$ is very close to 1, $T\sim-\frac{1}{2}\ln(1-k) \gg1$, and, thus, 
(\ref{final unstable man}) is not valid for $t>T$.
In this case, (\ref{final unstable man}) is applicable for $t\ll T$.
Indeed, when $k=1.0-1.4\times 10^{-12}$, (\ref{final unstable man}) shows self crossing curves.
Hence, Fig.~\ref{reconnection} is drawn by restriction of $t<T$.

\section{Conclusions}
With the aid of the ABAO (the asymptotics beyond all orders) method, we have 
derived analytical approximation of the unstable/stable manifolds, 
which agree rather well with those obtained by the numerical iteration.
The perturbed unstable manifold starting from $(\pi,0)$ 
acquires additional terms which are exponentially small with respect 
to perturbation parameter $\sigma$, but it exponentially grows with respect to 
time $t$.
Therefore separatrix splitting can be explained by the change of dominant terms
among perturbative solution and added terms derived by the ABAO method.
And overall approximation shows 
a highly oscillatory behavior only near the fixed point ($-\pi,0$) 
when $k$ is not close to 1.
From the comparison with the numerical calculation,
we check that the approximation breaks dowm when time 
goes to large and that the time regime where 
the approximation is valid extends thanks to the contribution from 
the higher order terms with respect to $\sigma$. 
We also mention that when $k$ is very close to 1
we can observe an oscillation near $(0,\pi)$, which is 
considered to be the precursor of the heteroclinic tangle in the unstable 
manifold at $k=1$.
In this way, even when the heteroclinic tangle exists, the unstable manifold smoothly
changes its topology as the change of the parameter $k$.

In the systems studied so far such as, the standard map, the H\'enon map, 
and the cubic map, the perturbative solution has only one sequence of 
singularities in the complex time domain.
On the contrary, the Harper map exhibits two sequences of singularities when $k\neq 1$.
The interference of the contribution from them might be expected. 
We concretely write down the interference term and show that 
it can be negligible.
Also we show that it is important, in general, to choose suitable 
initial condition because incorrect choice of the initial conditions 
lead to Stokes multipliers(SM) which diverge as $k\to 1$.
This also shows that even the perturbative solution 
does not have a symmetry with respect to $u$ axis.
In a higher order approximation, the interference should contribute 
and such a situation will be discussed elsewhere.

Before closing this section, we summarize the matching conditions and 
order of the added terms in the ABAO analysis for the following maps 
(table \ref{other}).\\
\noindent 
the standard map\cite{Gelfreich99,NakamuraKushibe}\ \ :
$
\pmat{v_{n+1} -v_n \cr u_{n+1}-u_n}=
\pmat{\sigma u_{n+1} \cr \sigma \sin v_n}
$
\\
\noindent
the H\'enon map\cite{Gelfreich4,Tovbis}\ \ \ \ \ :
$\pmat{v_{n+1} -v_n \cr u_{n+1}-u_n}=
\pmat{\sigma u_{n+1} \cr \sigma (v_n^2-2v_n)}$
\\ \noindent
the cubic map\cite{NakamuraHamada}\ \ \ \ \ \ \ \ \ \ \ :
$\pmat{v_{n+1} -v_n \cr u_{n+1}-u_n}=
\pmat{\sigma u_{n+1} \cr -4\sigma v_n(v_n-a)(v_n+a)}$ 
\\ \noindent
as well as the Harper map.

\begin{table}[h]
\begin{center}
\begin{tabularx}{137mm}{|l|l|l|X|}
\hline
~~~~Map & imaginary part of & matching between & order of \cr &a nearby singularity & 
leading order terms & added terms
 \\ \hline
Standard map 
& $t_c=\frac{i\pi}{2}$
&
~~~$v_{10}(t_p+\sigma z) \approx z^2 e^{\frac{2\pi i t_c}{\sigma}}$ &~~$\frac{1}{\sigma^2}e^{-\frac{\pi^2}{\sigma}}$
\\ \hline
H\'enon map & $t_c=\frac{i\pi}{\sqrt{2}}$
& $\sigma^2 v_{10}(t_p+\sigma z)  \approx z^4
e^{\frac{2\pi i t_c}{\sigma}}$
&
~~$\frac{1}{\sigma^6}e^{-\frac{\sqrt{2}\pi^2}{\sigma}}$  \\ \hline
Cubic map & $t_c=\frac{i\pi}{4a}$
& ~$\sigma v_{10}(t_p+\sigma z)  \approx z^3 e^{\frac{2\pi i t_c}{\sigma}}$
&
~~$\frac{1}{\sigma^4}e^{-\frac{\pi^2}{2\sigma a}}$
\\ \hline
Harper map &$t_c=\frac{i\pi}{2\sqrt{k}}
$
&~~~$v_{10}(t_p+\sigma z)  \approx z e^{\frac{2\pi i t_c}{\sigma}}$
&
~~$\frac{1}{\sigma}e^{-\frac{\pi^2}{\sigma\sqrt{k}}}$ \\ \hline
\end{tabularx}
\end{center}
\caption{Order of added terms and matching conditions for several maps.
The parameter $\sigma$ plays a role of the small parameter, 
$t_p$ stands for the singularity in the upper half plane nearest to the real axis and $t_c=t_p-{\rm Re}t_p$ and
$v_{10}(t)$ is the additional terms to the outer equation which appears when $t$ exceeds Re~$t_p$.}
\label{other}
\end{table}

\section*{Acknowledgments}
The authors thank Hidekazu Tanaka and  Satoshi Nakayama for their contributions
to the early stage of this work. Also, they are grateful to Prof. K. Nakamura, 
Prof. A. Shudo, Dr.S. Shinohara, Dr. T. Miyaguchi and Dr. G. Kimura for 
fruitful discussions and useful comments. 
One of them also thank Prof. D. Sauzin for fruitful discussions and encouragement.
Particularly, they thank Prof. Nakamura for turning their attention to 
Ref.\cite{NakamuraKushibe}. This work is partially supported by 
Grant-in-Aid for Scientific 
Research (C) (No.17540365) 
from the Japan Society of the Promotion of Science, 
by a Grant-in-Aid for Scientific Research of Priority Areas 
``Control of Molecules in Intense Laser Fields'' (No.14077219), 
the 21st Century COE Program at Waseda University ``Holistic Research 
and Education Center for Physics of Self-organization Systems''
and ``Academic Frontier'' Project 
from the Ministry of Education, Culture, Sports, Science and 
Technology of Japan, as well as by
Waseda University Grant for Special Research
Projects, Individual Research (2004A-161).

\appendix

\section{Relation between cases of $k>1$ and $0<k<1$}\label{appC}
In this section, we consider the case of $k>1$.
Put
\begin{eqnarray}
\tau\equiv kt,\ \widetilde{\sigma}\equiv k\sigma ,
\quad
\widetilde{u}(\tau)\equiv v\left(-\frac{\tau}{k}+\sigma\right),\ 
\widetilde{v}(\tau)\equiv u\left(-\frac{\tau}{k}+\sigma\right)
\label{trans}
\end{eqnarray}
then $\widetilde{v}$, $\widetilde{u}$ are the solution of the 
Harper map with parameter $1/k$:
\begin{eqnarray}
\widetilde{v}(\tau+\widetilde{\sigma})-\widetilde{v}(\tau)&=&
-\widetilde{\sigma} \sin \widetilde{u}(\tau)
\nonumber
\\
\widetilde{u}(\tau+\widetilde{\sigma})-\widetilde{u}(\tau)&=&
{\widetilde{\sigma}\over k} \sin \widetilde{v}(\tau+\widetilde{\sigma})
\end{eqnarray}

\section{Contribution from complex conjugate singularity}
\label{app conjugate}
In the main part of this article, we present details of analytic continuation 
via the behavior of the solution near $t_1$.
The purpose of this appendix is to prove that the additional term arrising from
$t_1^*$ is just the complex conjugate of that from $t_1$.
We present the proof for dominant terms, but we only use 
a real analyticity of perturbative solution.
Therefore, the same discussion can be used 
for less singular terms.
From a real analyticity of $(v_{0n},u_{0n})$ and the expansion
(\ref{for appen}) in a main part,  $(v_{0n},u_{0n})$ has a 
following expansions near $t_1$ and $t_1^*$ in the complex time domain.
\begin{eqnarray*}
\begin{pmatrix}
v_{0i}(t)\cr u_{0i}(t)
\end{pmatrix}
=\sum_{l=0}^\infty
\begin{pmatrix}
a_l^{(i)}(t-t_1)^l \cr b_l^{(i)}(t-t_1)^l
\end{pmatrix}
+\sum_{l=0}^\infty
\begin{pmatrix}
\overline{a_l^{(i)}}(t-t_1^*)^l \cr \overline{b_l^{(i)}}(t-t_1^*)^l
\end{pmatrix}
\ , 
\quad
(i\ge 1, b_0^{(i)}\not= 0) \ .
\end{eqnarray*}
Hence, this expansion gives the following asymptotic expansion:
\begin{eqnarray}
\sum_{i=1}^\infty {1\over z^{i}}
\pmat{a_0^{(i)}\cr b_0^{(i)}}
+\sum_{i=1}^\infty {1\over \widetilde{z}^{i}}
\pmat{\overline{a_0^{(i)}} \cr \overline{b_0^{(i)}}}
\label{both ex}
\end{eqnarray}
where $z$ and $\widetilde{z}$ are defined by $z=(t-t_1)/\sigma$ and $z=(t-t_1^*)/\sigma$.
Therefore, it is sufficient to show that the second term generates just 
the complex conjugate of the first term:
\begin{eqnarray*}
\pmat{V_{add}(z) \cr U_{add}(z)}\equiv
-\int_{\gamma} dp e^{-pz}
\sum_{i=1	}^\infty \frac{p^{i-1}}{(i-1)!}
\pmat{a_0^{(i)}\cr b_0^{(i)}}
\label{samp}
\end{eqnarray*}
Let $\gamma '$ denote symmetric pass of $\gamma$ with respect to real axis 
in $p$ plane.By analitically continuing the second term of (\ref{both ex}) from 
Re$\widetilde{z}<0$, Im$\widetilde{z}>0$ to 
Re$\widetilde{z}>0$, Im$\widetilde{z}>0$ 
following term is added.
\begin{eqnarray*}
-\int_{\gamma '} dp e^{-p\widetilde{z}}
\sum_{i=1}^\infty \frac{p^{i-1}}{(i-1)!}
\pmat{\overline{a_0^{(i)}} \cr \overline{b_0^{(i)}}}
&=&
-\int_{\gamma} d\bar{p} e^{-\overline{p}\widetilde{z}}
\sum_{i=1}^\infty \frac{\overline{p}^{i-1}}{(i-1)!}
\pmat{\overline{a_0^{(i)}} \cr \overline{b_0^{(i)}}}
\\&=&
-\int_{\gamma} dp e^{-p\widetilde{z}^*}
\sum_{i=1}^\infty \frac{p^{i-1}}{(i-1)!}
\pmat{a_0^{(i)} \cr b_0^{(i)}}
\\&=&
\pmat{\overline{V_{add}(\widetilde{z}^*)} \cr 
\overline{U_{add}(\widetilde{z}^*)}}
\end{eqnarray*}
near $t_1^*$.
Let us suppose that ${\displaystyle
x_{add}(t,\sigma)}$
is matched with (\ref{samp}) near $t_1$.
Ofcourse, the second term is matched with ${\displaystyle
\overline{ x_{add}(\overline{t},\sigma)} }$ and it 
is the complex conjugate of the first term, 
${\displaystyle x_{add}(t,\sigma)}$, 
for real $t$ and this fact complete the proof.

\section{Borel-transformed linearized inner equation \label{linearized}}
The aim of this appendix is to examine an asymptotic behavior of 
$(V_A(z),\ U_A(z))$ and $(V_B(z),\ U_B(z))$.
The Borel transforms $(\widetilde{V}_{A}(p),\widetilde{U}_{A}(p))$
of $(V_{A}(z),\ U_{A}(z))$ 
and $(\widetilde{V}_{B}(p),\widetilde{U}_{B}(p))$
of $(V_{B}(z)-z,U_{B}(z)-(z+\frac{1}{2}))$ satisfy

\begin{eqnarray}
(e^{-p}-1)\widetilde{V}_{A}(p)&=&\widetilde{U}_{A}(p)*g(p)
\nonumber
\\
(1-e^{p})\widetilde{U}_{A}(p)&=&\widetilde{V}_{A}(p)*f(p)
\nonumber
\\
(e^{-p}-1)\widetilde{V}_{B}(p)&=&\widetilde{U}_{B}(p)*g(p)+\frac{1}{2}g(p)+g'(p)
\nonumber
\\
(1-e^{p})\widetilde{U}_{B}(p)&=&\widetilde{V}_{B}(p)*f(p)+f'(p)
\label{linearBorel}
\end{eqnarray}
where
$$
B\left[\frac{e^{iV_{00}(z)}}{z}\right]=f(p)
,\ B\left[\frac{e^{- iU_{00}(z)}}{z}\right]=g(p)
$$
By substituting the power series expansions:
\begin{eqnarray}
\widetilde{V}_{A}(p)&\equiv&\sig{n}A^V_{n}p^{n}
, \quad
\widetilde{U}_{A}(p)\equiv\sig{n}A^U_{n}p^{n}
; \quad A_0^V=-1 , \quad A_0^U=1
\nonumber\\
\widetilde{V}_{B}(p)&\equiv&\sig{n}B^U_{n}p^{n}
, \quad
\widetilde{U}_{B}(p)\equiv\sig{n}B^U_{n}p^{n}
; \quad B_0^V=0 , \quad B_0^U=\frac{1}{12}
\nonumber
\end{eqnarray}
into (\ref{linearBorel}) and comparing term by term, the coefficients 
$a_{n},\ b_{n}$ are determined recursively and one has
\begin{eqnarray}
A^V_{2n-1}&=&0,\ \ \ \ \ \ \ \ (\forall n)
\nonumber\\
A^V_{2n}(-1)^{n}(2\pi)^{2n} &\rightarrow& -B_4 n
\quad (n\to \infty)
\nonumber\\
A^U_{2n-1}(-1)^{n+1}(2\pi)^{2n-1} &\rightarrow& -B_5
\quad (n\to \infty)
\nonumber
\\
A^U_{2n}(-1)^{n}(2\pi)^{2n} &\rightarrow& -B_4 n
\quad (n\to \infty)
\\
B^V_{2n-1}&=&0,\ \ \ \ \ \ \ \ (\forall n)
\nonumber\\
B^V_{2n}(-1)^{n}(2\pi)^{2n} &\rightarrow&B_1 (2n+2)(2n+1)-B_2 n
\quad (n\to \infty)
\nonumber\\
B^U_{2n-1}(-1)^{n+1}(2\pi)^{2n-1} &\rightarrow& -B_3 n
\quad (n\to \infty)
\nonumber
\\
B^U_{2n}(-1)^{n}(2\pi)^{2n} &\rightarrow& -B_1 (2n+2)(2n+1)-B_2 n
\quad (n\to \infty)
\end{eqnarray}

where $B_1=0.01480,\  \ B_2=0.14,\ B_3=0.186,\ B_4=3.503,\ B_5=5.551$.
Therefore, the following relations hold:
\begin{eqnarray*}
\pmat{\widetilde{V}_A(p) \cr \widetilde{U}_A(p)}&=&
i\pmat{2\pi^3 B_4 f_1^{(R)}(p)
\cr 2\pi^3 B_4 f_1^{(R)}(p)}
\\
\pmat{\widetilde{V}_B(p) \cr \widetilde{U}_B(p)}&=&i\pmat{8\pi^4 B_1 f_2^{(I)}(p)+2\pi^3B_2 f_1^{(R)}(p)
\cr 
-8\pi^4 B_1 f_2^{(I)}(p)+2\pi^3 B_2 f_1^{(R)}(p)}
+2\pi^3 i B_3 f_1^{(I)}(p)\pmat{0 \cr 1}
\end{eqnarray*}

\begin{eqnarray*}
-2\pi i e^{2\pi iz}\rm{Res}_{p=2\pi i} 
\pmat{\widetilde{V}_A(p)e^{-pz} \cr \widetilde{U}_A(p)e^{-pz}}&=&
2\pi^3 i B_4 z 
\pmat{1 \cr 1}
\\
-2\pi i e^{2\pi iz}\rm{Res}_{p=2\pi i} 
 \pmat{\widetilde{V}_B(p)e^{-pz} \cr \widetilde{U}_B(p)e^{-pz}}&=&
8\pi^4 B_1 z^2\pmat{1 \cr -1}
+2\pi^3 i B_2 z\pmat{1 \cr 1}
-2\pi^3 B_3 z\pmat{0 \cr 1}
\end{eqnarray*}

\section{Borel-transformed first order solution of inner equation \label{app01}}The aim of this appendix is to examine an asymptotic behavior of the solutions $V_{01}(z),\ U_{01}(z)$.
The Borel transforms $\widetilde{V}_{1}(p),\widetilde{U}_{1}(p)$
of $V_{01}(z),U_{01}(z)-i\frac{k-1}{4}$ satisfy 
\begin{eqnarray}
(e^{-p}-1)\widetilde{V}_{1}(p)&=&i\frac{k-1}{2}
g'(p)+i\frac{k-1}{4}g(p)+\widetilde{U}_{1}(p)*g(p)
\nonumber
\\
(1-e^{p})\widetilde{U}_{1}(p)&=&i\frac{k-1}{2}f'(p)+\widetilde{V}_{1}(p)*f(p)
\label{inn}
\end{eqnarray}

The power series expansions of $\widetilde{V}_1(p),\ \widetilde{U}_1(p)$ is defined by
\begin{eqnarray}
\widetilde{V}_{1}(p)=-i\frac{kt_1 +1}{24}+\sum_{n=1}^{\infty}\widetilde{c}_{n} p^{n}
\ , \qquad 
\widetilde{U}_{1}(p)=i\frac{k(t_1 +1)}{24}+\sum_{n=1}^{\infty}\widetilde{d}_{n} p^{n}
\label{exp}
\end{eqnarray}
(\ref{inn}) and (\ref{exp}) gives the following form:
\begin{eqnarray}
\widetilde{V}_1(p)=kt_{1}\widetilde{V}_x(p)+(k-1)\widetilde{V}_y(p)
+\widetilde{V}_z(p)
\ , \quad
\widetilde{U}_1(p)=kt_{1}\widetilde{U}_x(p)+
(k-1)\widetilde{U}_y(p)+\widetilde{U}_z(p)
\nonumber \\
\label{ex}
\end{eqnarray}
where $\widetilde{V}_{\alpha}(p),\ \widetilde{U}_{\alpha}(p)\ (\alpha=x,y,z)$ are 
independent of $k$.
By substituting (\ref{ex}) into (\ref{inn}), we get
\begin{eqnarray}
(e^{-p}-1)\widetilde{V}_{\alpha}(p)&=&\widetilde{U}_{\alpha}(p)*g
\nonumber
\\
(1-e^{p})\widetilde{U}_{\alpha}(p)&=&\widetilde{V}_{\alpha}(p)*f
\nonumber
\\
&&(\alpha=x,z)
\nonumber
\\
(e^{-p}-1)\widetilde{V}_{y}(p)&=&\widetilde{U}_{y}(p)*g+\frac{1}{4}g(p)+\frac{1}{2}g'(p)
\nonumber
\\
(1-e^{p})\widetilde{U}_{y}(p)&=&\widetilde{V}_{y}(p)*f+\frac{1}{2}f'(p)
\label{each}
\end{eqnarray}
By comparing (\ref{each}) with (\ref{linearBorel}), we get
\begin{eqnarray}
\widetilde{V}_{1}(p)=i\left(
\frac{kt_1+1}{24}\widetilde{V}_A(p)+\frac{k-1}{2}\widetilde{V}_B(p)\right)
\nonumber\\
\widetilde{U}_{1}(p)=i\left(
\frac{kt_1+1}{24}U_A(p)+\frac{k-1}{2}\widetilde{U}_B(p)
\right)
\label{couter integral ap}
\end{eqnarray}
This estimation gives (\ref{01appe}).

\section{Choice of initial time and Stokes Multiplier}\label{app i.c.}
In this appendix, the relation between initial time and Stokes multiplier
is discussed.
Here we restrict our discussion to the case of 
$(v_{02}(t),u_{02})$, which is used to obtain our final formula.
We start from pertarbative solution of $(v_{02}(t),u_{02}(t))$.
\begin{eqnarray*}
v_{02}(t)&=&
-\frac{1}{24}\left[
x_{1}'(t)+x_{1}(t)\left\{kt-2\sqrt{k}
\frac{\left(\sinh\sqrt{k}t - \sqrt{k}\cosh\sqrt{k}t\right)
\left(\cosh\sqrt{k}t - \sqrt{k}\sinh\sqrt{k}t\right)}
{(1+k)\cosh^2 \sqrt{k}t-2\sqrt{k}\sinh\sqrt{k}t\cosh\sqrt{k}t}
\right\}
\right]
\nonumber
\\
&+&\frac{A(k)}{24}x_1(t)+\frac{B(k)}{24}x_2(t)
\nonumber
\\
u_{02}(t)&=&
\frac{1}{24}\left[
2y_{1}'(t)-y_{1}(t)\left\{kt-2\sqrt{k}
\frac{\left(\sinh\sqrt{k}t - \sqrt{k}\cosh\sqrt{k}t\right)
\left(\cosh\sqrt{k}t - \sqrt{k}\sinh\sqrt{k}t\right)}
{(1+k)\cosh^2 \sqrt{k}t-2\sqrt{k}\sinh\sqrt{k}t\cosh\sqrt{k}t}
\right\}
\right]
\nonumber
\\
&+&\frac{A(k)}{24}y_1(t)+\frac{B(k)}{24}y_2(t)
\end{eqnarray*}
We note that $(A,B)$ is chosen as $(A,B)=(0,0)$ in the main part.
In this appendix, we put $B=0$ for $v_{02}(-\infty)=0,\ u_{02}(-\infty)=0$ 
and leave $A$.
Then, the set of second divergent terms in perturabative soluion near 
$t_1$ and $t_2$ are respectively
\begin{eqnarray}
\widetilde{V}_{1}(p)=i\left(
\frac{i\sqrt{k}\pi-2A(k)+2}{48}\widetilde{V}_A(p)+\frac{k-1}{2}\widetilde{V}_B(p)\right)
\nonumber\\
\widetilde{U}_{1}(p)=i\left(
\frac{i\sqrt{k}\pi-2A(k)+2}{48}U_A(p)+\frac{k-1}{2}\widetilde{U}_B(p)
\right)
\end{eqnarray}
and
\begin{eqnarray}
\widetilde{V}_{1}(p)=i\left(
\frac{4kT+i\sqrt{k}\pi-2A(k)-2}{48}
\widetilde{V}_A(p)+\frac{k-1}{2}\widetilde{V}_B(p)\right)
\nonumber\\
\widetilde{U}_{1}(p)=i\left(
\frac{-4kT-i\sqrt{k}\pi+2A(k)+2}{48}U_A(p)+\frac{k-1}{2}\widetilde{U}_B(p)
\right)
\end{eqnarray}
They derive the following Stokes multipliers.
\begin{eqnarray}
\Lambda^B_{1}&=&-(k-1)\pi^3 B_2-\frac{i\sqrt{k}\pi-2A(k) +2}{24}\pi^3 B_4
\nonumber
\\
\Lambda^B_{1}&=&-(k-1)\pi^3 B_2+\frac{4kT+i\sqrt{k}\pi-2A(k) -2}{24}\pi^3 B_4
\label{StokesValST}
\end{eqnarray}
where the upper multiplier is for $t_1$ and the lower one is for $t_2$.
We note that the choice $A(k)=kT=\sqrt{k}\ln{1+\sqrt{k}\over \sqrt{1-k}}$ makes a symmetric perturbative solution with 
respect to $u$ axis but it also makes the upper multiplier divergent.
Therefore this choice should be rejected.

\section{Matching of higher order Terms}\label{GeneralMatch}
In this subsection, Matching between inner and outer equation
for higher orders is discussed.
The analytic continuation of a inner solution $V_{0l}$ to the next
sector is:
$$
V_{0l}(z)+e^{-2\pi i z}V_{1l}(z)
$$$$
U_{0l}(z)+e^{-2\pi i z}U_{1l}(z)
$$
where $V_{1l},\ U_{1l}$ should be determined by the linearized equation.
To deal with more general case, we analyze the following 
linearized inner equation:
\begin{eqnarray}
\Delta  V_1(z,\sigma)=
\sig{n}\left\{
f_{1n}(z)\sigma^n V_1 (z,\sigma)+f_{2n}(z)\sigma^n U_1 (z,\sigma)
\right\}
\nonumber\\
\Delta U_1 (z,\sigma)=
\sig{n}\left\{
g_{1n}(z)\sigma^n V_1 (z,\sigma)+g_{2n}(z)\sigma^n U_1 (z,\sigma)
\right\}
\label{2ndlinear}
\end{eqnarray}
Next, we seek to find the solution of the type
\begin{eqnarray}
\pmat{V_1(z) \cr U_1(z)}=\sig{l}\sigma^l\pmat{V_{1l}(z) \cr U_{1l}(z)}
\end{eqnarray}
Because (\ref{2ndlinear}) is linear and second difference equation, each 
$\pmat{V_l(z) \cr U_l(z)}$ has two undefined constant 
corresponding to homogeneous solution. 
It indicates that $V_1,\ U_1$ has the form:
\begin{eqnarray}
\pmat{V_{1}(z) \cr U_{1}(z)}&=&
\sig{l}\sigma^l\pmat{V_{1l}(z) \cr U_{1l}(z)}
\nonumber\\
&=&\sum_{\alpha}\sig{i}
\Lambda_\alpha(\sigma)
\pmat{\sigma^{i}V^\alpha_{1i}(z) \cr \sigma^{i}U^\alpha_{1i}(z)}
\label{general in}
\end{eqnarray}
where $\Lambda_\alpha(\sigma)$ can be expanded as a power series of $\sigma$ 
and where $\pmat{V_{\alpha}(z) \cr U_{\alpha}(z)}$, 
$\pmat{V^{\alpha}_{1i}(z) \cr U^{\alpha}_{1i}(z)}$
are defined by the following rules.
\begin{enumerate}
\item
$\pmat{V_{\alpha}(z) \cr U_{\alpha}(z)}$ is the independent solutions of $\pmat{V_{10}(z) \cr U_{10}(z)}$ 
which satisfy
$$\lim_{|z|\rightarrow\infty}z^{-j_\alpha}V_{\alpha}(z)=1\ (\alpha=A,B),\ 
j_A<j_B$$
Note that this condition is not imposed on $U_\alpha$.
\item
$\pmat{V^\alpha_{1l}(z) \cr U^\alpha_{1l}(z)}$ is the solution when 
$\pmat{V_{10}(z) \cr U_{10}(z)}=\pmat{V_\alpha(z) \cr U_\alpha(z)}$ and 
$\pmat{V_{1i}(z) \cr U_{1i}(z)} =\pmat{V^\alpha_{1i}(z) \cr U^\alpha_{1i}(z)}$
$(i\ge l-1)$. More concretely, they are inductively defined by 
the following equations.
$$
\Delta\pmat{V^{\alpha}_j(z) \cr U^{\alpha}_j(z)}
=\sum_{j=0}^{l}
\pmat{f_{1 j}(z)V^{\alpha}_{l-j}(z)+f_{2 j}(z)U^{\alpha}_{l-j}(z)
\cr
g_{1 j}(z)V^{\alpha}_{l-j}(z)+g_{2 j}(z)U^{\alpha}_{l-j}(z)
}$$

\item
\{coefficient of $z^{j_A}$ and $z^{j_B}$ in $V^\alpha_{1l}(z)$\} =0
\end{enumerate}
In (\ref{2.2st}), boundary conditions for $v_{ni}\ ,\ u_{ni}\ (n\ge 1)$ 
and $\epsilon$ are unsettled.
On the other hands, $\Lambda_l^A,\ \Lambda_l^B$ in (\ref{general in}) 
is determined by analytic continuation of asymptotic expansions.
Therefore $\Lambda_l^A,\ \Lambda_l^B$ determine (\ref{2.2st})
by matching the asymptotic expansions with it.
Similar to the case of $(V^\alpha_1,U^\alpha_1)$,
solutions of $v_{1l}(t),\ v_{1l}(t)$ are
\begin{eqnarray*}
\pmat{v_{10} \cr u_{10}}
&=&c^A_0 \pmat{v_A(t) \cr u_A(t)}
+c^B_0 \pmat{v_B(t) \cr u_B(t)}
\\
\pmat{v_{1l}(t) \cr u_{1l}(t)}&=&\sum_{j=0}^{l}
\left[
c^A_j\pmat{v^A_{1\ l-j}(t) \cr u^A_{1\ l-j} (t)}+
c^B_j\pmat{v^B_{1\ l-j}(t) \cr u^B_{1\ l-j} (t)}
\right]
\nonumber\\.
\end{eqnarray*}
where $\pmat{v_\alpha \cr u_\alpha}$ denote the homogeneous solution 
which satisfy:
\begin{eqnarray*}
1=\lim_{t\rightarrow t_1}(t-t_1)^{-j_\alpha} v_\alpha(t)
\end{eqnarray*}

and where $v^\alpha_{1l}(t),\ u^\alpha_{1l}(t),\ (l\ge 1)$
are defined by following 2 conditions.
\begin{enumerate}
\item{$\pmat{v^\alpha_{1l} \cr u^\alpha_{1l}}$ is the solution when 
$\pmat{v_{10} \cr u_{10}}=\pmat{v_\alpha \cr u_\alpha}$ and 
$\pmat{v_{1i} \cr u_{1i}} =\pmat{v^\alpha_{1i} \cr u^\alpha_{1i}}
(0<i\le l-1)$}
\item{\{coefficient of $(t-t_1)^{j_\alpha}$ 
in $v^\alpha_{1l}$\} =0}
\end{enumerate}
The matching condition is:
\begin{eqnarray}
e^{-2\pi iz}\sig{l}\sigma^l V_{1l}(z)
=E(t)\sig{i}\sigma^iv_{1i}(t)
\label{matchingintro}
\end{eqnarray}
As a result, we get
\begin{eqnarray*}
E(t)&=&\sigma^{-j_B}
e^{-2\pi iz}
\nonumber
\\
\sig{l}\sigma^l \pmat{V_{1l}(z) \cr U_{1l}(z)}
&=&\sigma^{-j_B}\sig{i}\sigma^i
\pmat{v_{1i}(t) \cr u_{1i}(t)}
\end{eqnarray*}
and the second equality leads
\begin{eqnarray*}
c^A_j&=&0, (j<j_B-j_A)
\nonumber\\ 
c^A_j&=&\Lambda^A_{j-j_A-j_B},\ \ (j\ge j_B-j_A)
\nonumber\\ 
c^B_j&=&\Lambda^B_j .
\end{eqnarray*}
The rest of this appendix is devoted to present a concrete 
calculation for the Harper map.
The Harper map has the following values.
$$j_A=-1,\ j_B=1$$
It indicates the following relation.
\begin{eqnarray*}
E(t)=\frac{1}{\sigma}
e^{-2\pi iz}=\frac{1}{\sigma} e^{\frac{2\pi it_1}{\sigma}}
\nonumber
\\
\sig{l}\sigma^l \pmat{V_{1l}(z) \cr U_{1l}(z)}
=\frac{1}{\sigma}\sig{i}\sigma^i
\pmat{v_{1i}(t) \cr u_{1i}(t)}
\end{eqnarray*}

As noted above, the coefficient 
$\displaystyle \Lambda_\alpha(\sigma)\equiv\sig{n}\sigma_n \Lambda_n^\alpha$
determine whole solutions in (\ref{2.2st}).
Let us concretely write down the first two terms:
\begin{eqnarray*}
&&
\left[
\begin{array}{l}
\displaystyle -\int_{\gamma} dp \hspace*{1mm} e^{-pz} \hspace*{1mm}
\{\widetilde{V}_0(p) + \sigma \widetilde{V}_1(p)\}
\\
\displaystyle -\int_{\gamma} dp \hspace*{1mm} e^{-pz} \hspace*{1mm}
\{\widetilde{U}_0(p) + \sigma \widetilde{U}_1(p)\}
\end{array}
\right]
\nonumber
\\
\nonumber
\approx&&\left(
\Lambda_B
\left[
\begin{array}{l}
(V_{10}^{B}(z)+V_{11}^{B}(z)\sigma+\cdots)
\\ 
(U_{10}^{B}(z)+U^{B}_{11}(z)\sigma+\cdots)
\end{array}
\right]
+
\Lambda_A
\left[
\begin{array}{l}
(V_{10}^{A}(z)+V_{11}^{A}(z)\sigma+\cdots)
\\ 
(U_{10}^{A}(z)+U^{A}_{11}(z)\sigma+\cdots)
\end{array}
\right]
\right)
e^{-2\pi iz}
\nonumber
\\
\approx
&&\left[
\begin{array}{l}
\Lambda^B_{0}z+\sigma 
(\Lambda^B_{1}z + \frac{k-1}{6}\Lambda^B_{0}z^{2})
\\
\Lambda^B_{0}z+\sigma\left(
\Lambda^B_{1}z - \frac{k-1}{6}\Lambda^B_{0}(z^{2}+z)
\right)\end{array}
\right]
e^{-2\pi iz}
\end{eqnarray*}
where the contour integral is evaluated with the aid of 
(\ref{sol:numerical0}) and (\ref{sol:borel11ST}).
This leads to the following relations
\begin{eqnarray}
\Lambda^B_{0}&=&i4\pi^{3}A_{1}=24\pi^{4}iB_{1}=6\pi^{3}iB_{3}
\nonumber
\\
\Lambda^B_{1}&=&-(k-1)\pi^3 B_2-\frac{kt_1 +1}{12}\pi^3 B_4
\label{StokesValST}
\end{eqnarray}
The above relations imply two linear relations among
$A_1$, $B_1$ and $B_3$, which are satisfied by the present
numerical estimations rather well:
\begin{eqnarray*}
{A_1\over 6 \pi B_1}=0.99998\simeq 1 \qquad
{2A_1\over 3 B_3}=0.99997\simeq 1 \nonumber \ .
\end{eqnarray*}

\end{document}